\documentclass[twocolumn]{aastex61}
\usepackage{amsmath}

\received{July 1, 2016}
\revised{September 27, 2016}
\accepted{\today}
\submitjournal{AJ}

\newcommand{\changed}{} 

\shorttitle{Nautilus: Very Large Aperture, Ultralight Space Telescope Concept}
\shortauthors{Apai et al.}

\begin{document}

\title{A Thousand Earths: A Very Large Aperture, Ultralight Space Telescope Array for Atmospheric Biosignature Surveys}

\correspondingauthor{D\'aniel Apai}
\email{apai@arizona.edu}

\author[0000-0003-3714-5855]{D\'aniel Apai}
\affil{Steward Observatory and the Lunar and Planetary Laboratory, The University of Arizona, Tucson, AZ 85721}

\author{Tom D. Milster, Dae Wook Kim}
\affil{James C. Wyant College of Optical Sciences, The University of Arizona, Tucson, AZ 85721}

\author[0000-0003-2831-1890]{Alex Bixel} 
\affil{Steward Observatory, The University of Arizona, Tucson, AZ 85721}

\author[0000-0002-4511-5966]{Glenn Schneider} 
\affil{Steward Observatory, The University of Arizona, Tucson, AZ 85721}

\author{Ronguang Liang}
\affil{College of Optical Sciences, The University of Arizona, Tucson, AZ 85721}

\author[0000-0003-1096-5634]{Jonathan Arenberg}
\affil{Northrop Grumman Aerospace Systems, Redondo Beach, CA 90278}



\begin{abstract}
An outstanding, multi-disciplinary goal of modern science is the study of the diversity of potentially Earth-like planets and the search for life in them. This goal requires a bold new generation of space telescopes, but even the most ambitious designs yet hope to characterize several dozen potentially habitable planets. Such a sample may be too small to truly understand the complexity of exo-earths. We describe here a {\changed notional} concept for a novel space observatory designed to characterize 1,000 transiting exo-earth {\changed candidates}. The Nautilus concept is based on an array of inflatable spacecraft carrying very large diameter (8.5m), very low-weight, multi-order diffractive optical elements (MODE lenses) as light-collecting elements. The mirrors typical to current space telescopes are replaced by MODE lenses with a 10 times lighter areal density that are 100 times less sensitive to misalignments, enabling light-weight structure. MODE lenses can be cost-effectively replicated through molding. {\changed The Nautilus mission concept has a potential {\changed to greatly reduce} fabrication and launch costs, and mission risks} compared to the current space telescope paradigm through replicated components and identical, light-weight unit telescopes. {\changed Nautilus is designed to survey} transiting exo-earths for biosignatures up to a distance of 300 pc, enabling a rigorous statistical exploration of the frequency and properties of life-bearing planets and the diversity of exo-earths. 
\end{abstract}




\section{Introduction} \label{sec:intro}

One of the pivotal questions of modern science is whether life is common in the Universe. Answering this question will most likely require measuring the occurrence rate of habitable planets, understanding their diversity, sampling their atmospheres, and determining whether the observed atmospheric compositions can be explained without biological activity (see, e.g., \citealt{Seager2016,Fujii2018,Kiang2018}). Although the characterization of exo-earths is a key science goal of next-generation telescopes \citep[][]{LUVOIR2018,HabEx2018,OST2018}, due to the challenging nature of the observations most proposed telescope concepts may not be able to accomplish this goal on {\changed target samples large enough to allow statistical exploration in a multi-dimensional parameter space \citep[e.g.,][]{Ramirez2019}. }
 A challenge central to these observations is the intrinsic faintness of exoplanets, which is further complicated by the close angular proximity of their bright host stars (planet/star contrast).

\begin{figure*}[!htbp]
\epsscale{1.0}
\plotone{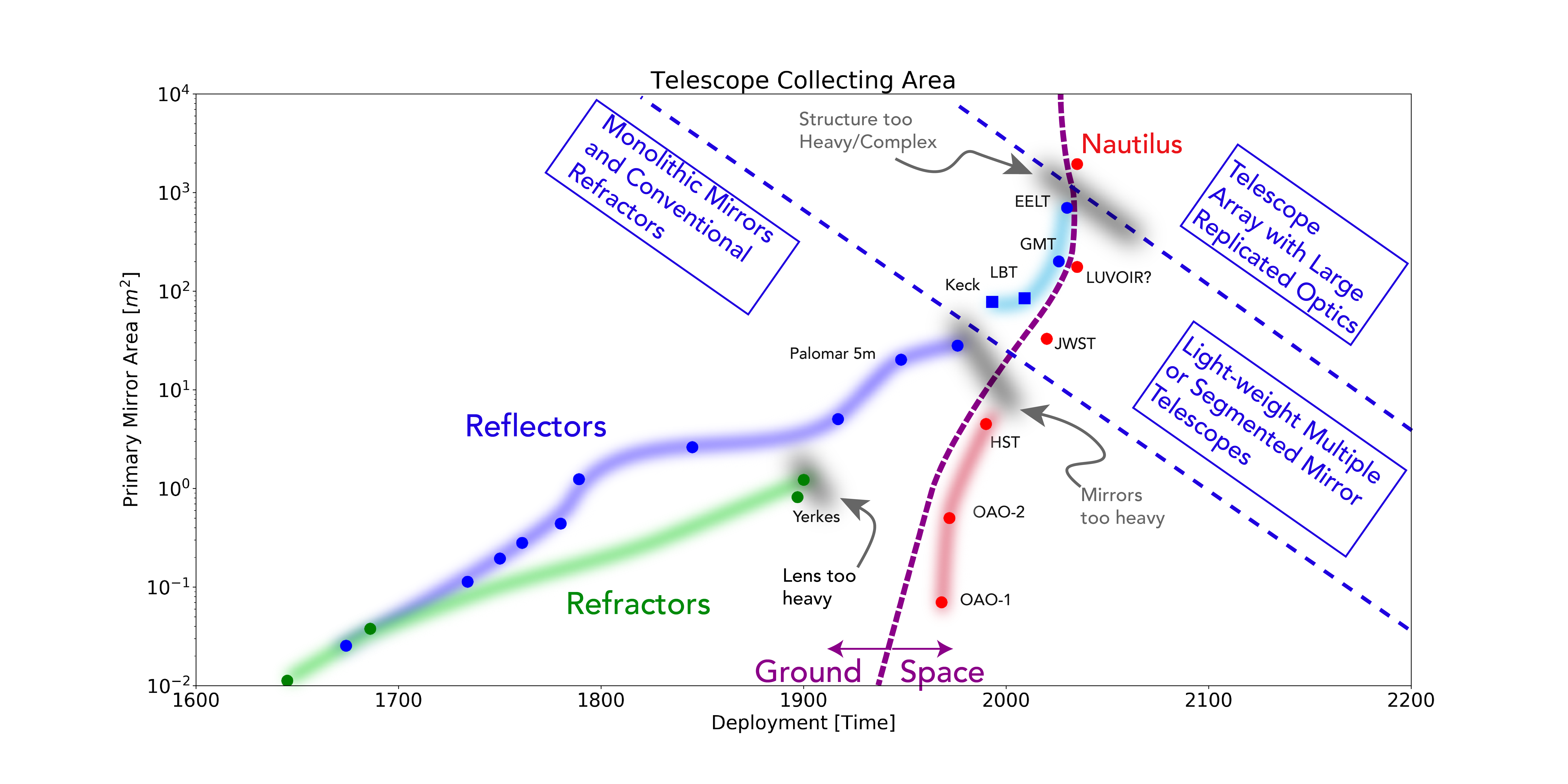}
\caption{The evolution of light-collecting area of ground-based (blue, green) and space-based (red) telescopes. The evolution is characterized by alternating stages of slow growth (when existing technology is scaleable) and pauses (when existing technology cannot be scaled up). 
The data points represent the installation of the largest telescopes in their era and are connected to highlight general trends. Gray regions mark the approximately stages in the evolution when lenses, monolithic mirrors, and then segmented mirrors become to massive to be viable with existing technology. Telescopes used multiple different technological solutions to collect light. Large jumps in diameter are due to change in technology due to technological breakthroughs. 
\label{PrimaryMirrors}}
\end{figure*}

One of the most fundamental properties of telescopes -- and a limiting factor for many studies of faint extrasolar planets -- is light-collecting area. For over a century -- following the commissioning of the 1.02m diameter Yerkes observatory refractor -- every large telescope built used a primary mirror to collect light -- but large mirrors (D$>$2.5~m) remained very difficult and expensive to fabricate, align, and operate. Figure~\ref{PrimaryMirrors} provides an overview of the evolution of light-collecting power and technology used over the past four centuries. While initially refracting and reflecting telescopes had been both utilized, reflectors proved to be scalable in size beyond refractors. Eventually, however, both refractors and reflectors reached the diameter beyond which their primary light-collecting elements (lenses and mirrors) became too heavy to maintain their figures. While functional refractors never exceeded 1.1m in diameter, functional monolithic mirrors could be built as large as 5~m. Manufacturing  large-diameter ($D$) mirrors with very high optical quality capable of working at optical wavelengths ($\lambda$) (D$>$6m and $\lambda\simeq0.5\mu$m) has been a technological challenge. 

For ground-based telescopes, after a four decade gap, the advent of segmented mirrors and ultralight (honeycomb) mirrors with computer-controlled surfaces enabled larger apertures. These technologies also enable the next generation of telescopes (Extremely Large Telescopes, ELTs) with effective diameters between 24.5m and 39.3m \citep[][]{Gilmozzi2007,Johns2012,Sanders2013}. It is unclear whether the same technology could be utilized to build 100~m-class ground-based telescopes.

In space, monolithic mirrors have been used for the largest visual/near-infrared astronomical telescopes, with the Hubble Space Telescope's (HST's) 2.4~m diameter mirror being the largest such element.
The James Webb Space Telescope ($D\sim6.5$~m) and some future concepts (such as the Large UV/Optical/IR Surveyor, LUVOIR) envision building on this heritage {\changed but utilizing segmented and actively controlled mirrors}.

While further slow, gradual increases in the diameter of segmented mirrors is possible, mirrors arguably remain the single most important bottleneck in astronomical telescopes.  With the slow growth of aperture sizes obvious next steps in astrophysics remain beyond reach: for example, the study of the diversity of Earth-like planets and assessing the frequency of atmospheric biosignatures in large samples ($N\sim1,000$) of Earth-like planets  remains beyond the reach of the telescopes envisioned even for the next forty years.
    
    We describe here a telescope concept that replaces the primary telescope mirror with multi-order diffractive engineered (MODE) material lens  technology \citep[][]{Milster2018PatentA,Milster2018PatentB}, offering a scalable solution for astronomical telescopes  with low production, launch, and alignment costs compared to modern reflecting telescopes. Our mission concept is called Nautilus, named after J. Verne's submarine. 
    Nautilus's {\em science goal} is to survey one thousand transiting, habitable zone Earth-sized exoplanets to determine their atmospheric diversity and the occurrence rate of atmospheric biosignatures. This planet sample represents 1--2 orders of magnitude increase over the direct imaging and exoplanet transit telescope concepts currently envisioned for the next three decades (e.g., LUVOIR concept: \citet{LUVOIR2018}, HabEx concept: \cite{HabEx2018}, OST concept: \cite{OST2018}).  
    
We will first review the physical principles behind  diffractive optical elements (\S\,\ref{IntroDOE}), design considerations (\S\,\ref{DesignConsiderations}) and the mass advantage (\S\,\ref{LensWeight}) they represent, then we summarize relevant past telescopes and telescope concepts based on large diffractive optics (\S\,\ref{PastProjects}).    
In \S~\ref{BiosignatureSearch} we review the key elements of a large-scale atmospheric biosignature survey for Earth-like exoplanets, including the methodology, sample size and definition  (\S\,\ref{S:SampleSize} and \S\,\ref{S:SampleDefinition}), followed by simulated results (\S\,\ref{S:SpecSim}). This is followed by a summary of the science requirements (\S\,\ref{SciReq}) for the survey and the baseline concept for the Nautilus Telescope Array (\S~\ref{S:NautilusArray}), including its launch, deployment, and operations. In \S\,\ref{S:Fabrication} we discuss the fabrication and scalability of MODE lenses, current prototypes, and real-time optical quality assessment considerations for fabrication of large-scale diffractive lenses. 
Finally, we discuss how the Nautilus concept compares to the state of the art, and discuss key advantages (reduced fabrication and launch costs, scalability) and the anticipated impact on astrophysics (\S\,\ref{S:Discussion}).

\subsection{Principles of Large-scale Diffractive Optical Elements}
\label{IntroDOE}

\begin{figure*}[!htbp]
\epsscale{1.2}
\plotone{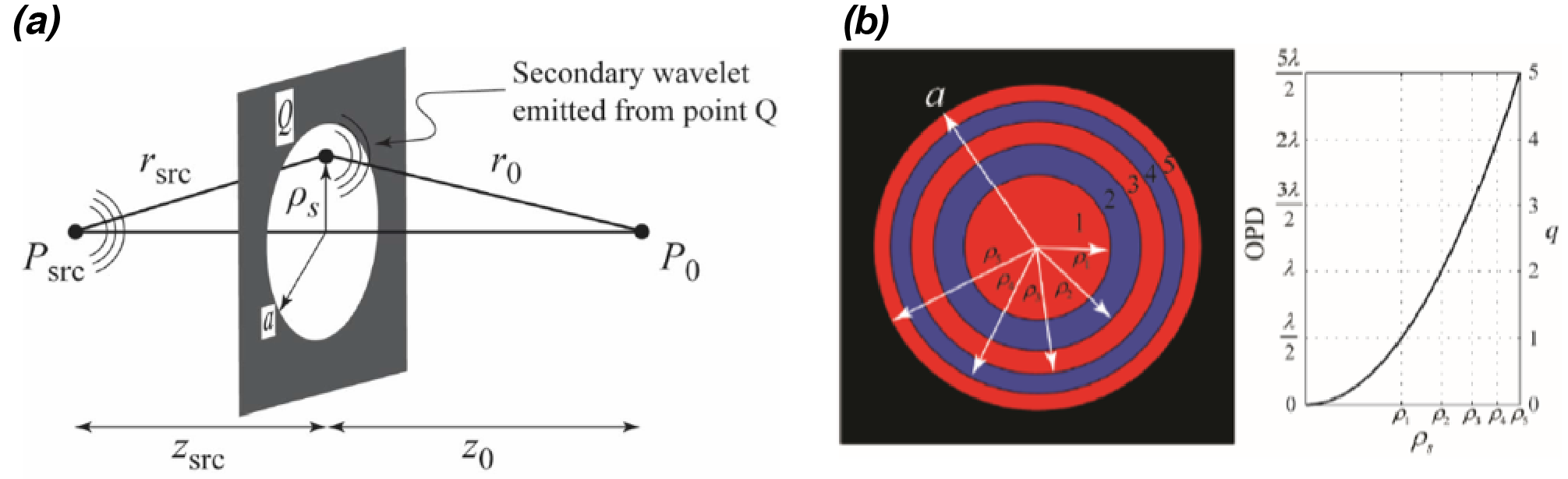}
\caption{Illustration of Fresnel zones and optical path difference (OPD). {\em (a)} Diffraction from an aperture illuminated by a point source is analyzed by considering changes in OPD as a function of radius in the aperture; and {\em (b)} Fresnel zone boundaries are defined by $\lambda/2$ changes in OPD. \label{FigTom1}}
\end{figure*}

Diffractive optical elements perform lens-like functions, which can be analyzed with the principle of interference. For example, light transmitted through an aperture with radius $a$ that is illuminated by point source $P_{src}$ is conceptually illustrated in Figure~\ref{FigTom1}a.  The aperture can be divided into equal-area Fresnel zones that identify which parts of the transmitted light interfere constructively at the observation point $P_0$ and which parts interfere destructively, as shown in Figure~\ref{FigTom1}b, based on the optical path difference (OPD) through point $Q$ of $(r_{src} + r_0) - (z_{src} + z_0)$.   Boundaries of the Fresnel zones are defined by an increase of $\lambda/2$ in OPD between successive zones.  In this example, the first and second Fresnel zones produce a net zero light amplitude at the observation point, because light from an even-numbered zone combines destructively with light from an odd-numbered zone due to the $\lambda/2$ OPD  between them.  Likewise, light from the third and fourth zones combine destructively, leaving only light from the fifth zone to produce non-zero light amplitude at the observation point.

The well-known Fresnel zone plate (FZP) operates by blocking only the even or odd zones in the aperture, thus producing only constructive wave combination at the observation point.  By extending this argument to off-axis illumination, it is understood that the FZP acts as a lens with a focus spot size that is equivalent to a classical lens of the same diameter and focal length.  However, due to the fact that other focal positions can be identified along the axis, the classical FZP results in high intensity background levels at the primary focus.  In addition, since the constructive or destructive nature of the wave combination depends on wavelength, the focal point changes chromatically with a focal length proportional to $1/\lambda$.  That is, as wavelength increases FZP focal length decreases, which is opposite the sense of a classical refractive lens \citep[][]{Milster2018Notes}.  The combination of a properly designed FZP on a refractive singlet leads to compensating focal dispersions, which results in an achromatic singlet \citep[][]{StoneGeorge1988}.

\begin{figure*}[!htbp]
\epsscale{0.7}
\plotone{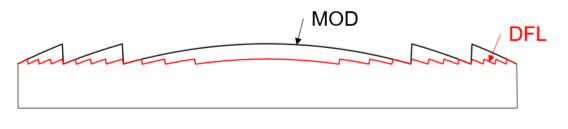}
\caption{Profiles of a diffractive Fresnel lens (DFL) and a multiple-order diffraction (MOD) lens.  The MOD lens is M times thicker and its zones are M times wider than the corresponding DFL. In this figure, $M = 4$. \label{FigTom2}}
\end{figure*}

In order to increase the diffraction efficiency of light into the desired primary focal order, the FZP is replaced by a diffractive Fresnel Lens (DFL), in which the opaque-zone FZP is replaced by a transmissive phase pattern that changes OPD as a function of radius.  Neighboring zones are combined into a single quadratic phase surface, as shown in Figure~\ref{FigTom2}.  The profile in each zone pair has a maximum of 1 wavelength of OPD across it.  Although the DFL has the same chromatic dispersion properties of a FZP, diffraction efficiency into the desired focal order is much greater.  In fact, under ideal conditions, all of the light is focused into the primary order.  Since the step height to achieve 1 wave of OPD at the transitions is very small (about 1 micron for visible light), the DFL is an extremely thin, planar optical element.

In order to decrease chromatic focal dispersion, a multiple-order diffractive (MOD) lens was developed \citep[][]{FaklisMorris1995}.  Instead of setting phase transitions based on a single wave of OPD, phase transitions are defined based on integer multiples of $M$ waves of OPD, where $M$ is the MOD number.  As shown in \ref{FigTom2} for $M = 4$, the MOD lens profile is thicker than the DFL by a factor of $M$ and zone spacings are increased.  However, even if $M = 1000$, the transition step is only about 1\,mm high for a visible light design.  The MOD lens operates over a set of higher diffracted orders where each order contains a wavelength of peak diffraction efficiency and each of these wavelengths come to a common focus. The lens exhibits strong chromatic dispersion at intermediate wavelengths but, interestingly, the maximum focal dispersion of the MOD lens is decreased to a range of approximately $f/M$ compared to the large range of a DFL.  For example, an $f = 5\,m$ focal length $M = 1,000$ MOD would have a focal range of only $\pm0.005\,m$ over a wavelength range from 500\,nm to 1000\,nm, where a DFL would have a focal range of approximately $\pm3.0m$ over the same range of wavelengths.

\subsection{Design Considerations for Multi-Order Diffractive Engineered Lenses}
\label{DesignConsiderations}
Design of a MODE lens for a particular application begins with the same desired first-order properties as a traditional refractive lens, such as the operating wavelengths and focal length. When designing for broadband performance the design wavelength is the central wavelength of the wavelength range. Transition depths are defined based on the formula $M {\lambda \over n_2-n_1}$  where $n_2$  and $n_1$ are the index of refraction of the lens material and the {incident index}, respectively. Transition locations are based on integer multiples of $M$ waves of OPD for on-axis rays. The individual zones are modelled and optimized in standard lens design software. 

The change of refractive index with wavelength in each radial zone of the MOD surface also produces a chromatic focal shift, although it is much smaller in magnitude than for a DFL. This problem is compensated for by fabricating a weak DFL on the rear surface of the MOD lens to create an achromatic singlet for each MOD radial zone. This combination is what we call a MODE lens \citep[][]{Milster2018PatentA,Milster2018PatentB}. The DFL is incorporated into the design by using a Sweatt model surface \citep[][]{Sweatt1977} in lens design software, in which a fictitious glass with index approximately equal to the wavelength in nm is used to allow for significant optical power to exist in a very thin region. The Sweatt surface must later be converted to a physical surface by wrapping the OPD it produces. The MOD surface cannot be modelled as a Sweatt surface as it no longer has a negligible physical thickness. 

Following optimization in lens design software, the MODE design is verified using a physical optics simulation to confirm the diffractive performance. Optical path length is determined at the exit pupil reference sphere using ray tracing. A Hankel transform calculation is used to determine field values at a sampled image plane. The magnitude squared of these field values provides the irradiance which represents the point spread function of the lens. This simulation is performed over a finely sampled spectrum of the full bandwidth as well as for a range of image planes to account for both refractive and diffractive chromatic dispersion.

\subsection{Mass Advantages of MODE Lenses}
\label{LensWeight}

In this section we briefly explore the mass reduction achieved by MODE lenses compared to conventional lenses. The discussion presented here aims to provide a first, general approximation, from which individually-designed MODE lenses may differ slightly.

We describe the mass of a conventional (refractive) lens as a planoconvex lens (sphere cap) as:

\begin{equation}
 M_{lens} =  \rho {1 \over 6} \pi h (3 R^2 + h^2),
\end{equation}

where $\rho$ is the density of the glass, $h$ is the height of the lens, and $R$ is the radius of the lens. 
In contrast, the mass of a MODE lens (of order $M$) designed for a center wavelength $\lambda_c$ is simply:

\begin{equation}
 M_{mode} =  \phi \rho (R^2 \pi) (M \lambda_c),
\end{equation}

where $\phi$ is the volume-filling factor of the MODE lens, close to 0.5. 

The mass ratio of a refractive lens to a MODE lens is then:
\begin{equation}
{M_{lens} \over M_{mode}} = {h (3 R^2 + h^2)  \over 6 \phi R^2 M \lambda_c} 
\end{equation}

For cases of relatively thin lenses ($h<R$) this ratio can be approximated to the first order by:

\begin{equation}
{M_{lens} \over M_{mode}}\approx { h \over M \lambda_c }.
\end{equation}

For example, for a lens with a radius of 5m and a relative thickness of h/R=0.1,  an M=1,000 MODE lens optimized for $\lambda_c=600$~nm would provide about two orders of magnitude of mass reduction. The mass reduction by replacing a thick lens with a MODE lens would be even greater. In short, MODE lenses represent at least two orders of magnitude lower mass for a given lens diameter, a transformative advantage for space telescopes.

\subsection{Telescopes based on Diffractive Optical Elements}
\label{PastProjects}

Diffractive optical elements (DOEs) are used for both space and commercial applications as small-scale internal optics. For example, the Lunar Orbiter Laser Altimeter (LOLA) mission uses a glass DOE as a beam splitter to divide a laser beam \citep[][]{Ramos-Izquierdo2009}. The high-quality Canon EF telephoto zoom lenses incorporate two diffractive optical surfaces that are first diamond turned in a mold and then replicated onto a curved glass substrate with epoxy resins \citep[][]{Nakai2003}. {\changed Similarly, Nikon's new {\changed Phase Fresnel (PF)} high-end telephoto lens series utilizes a Fresnel lens in combination with refractive {\changed } group to allow large-aperture, high-{\changed q}uality, but very light photolenses.}

\begin{figure*}[!htbp]
\epsscale{1.12}
\plotone{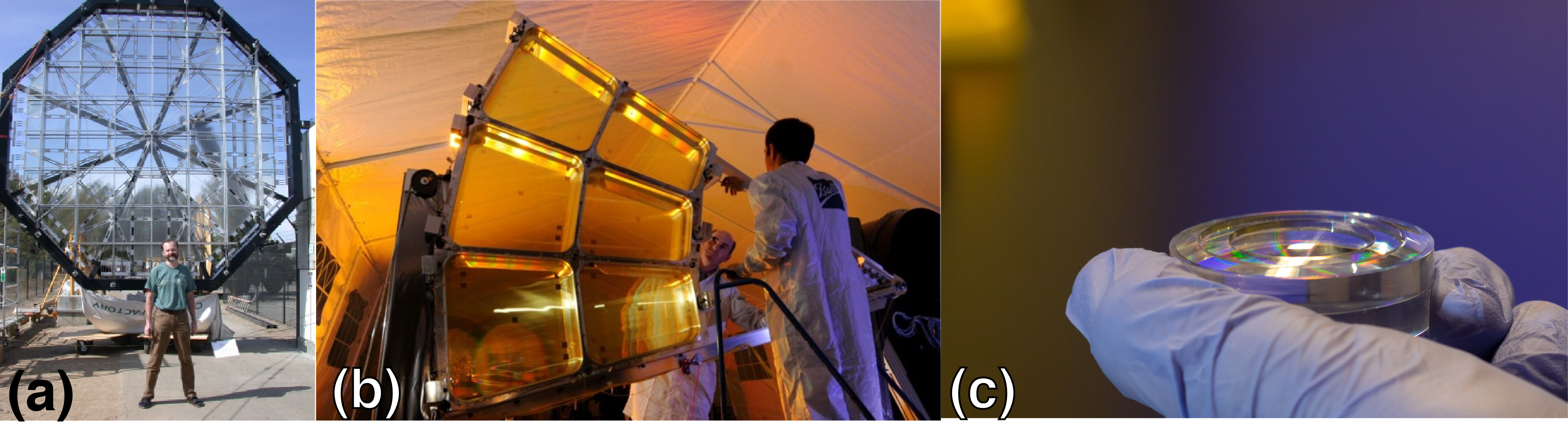}
\caption{a) 5m diameter Eyeglass prototype using 72 single-order diffractive optical elements \citep[e.g.,][]{Hyde1999}. b) Ball's MOIRE test segment, using single-order diffractive optics replicated to membranes \citep[][]{Atcheson2014}. c) MODE lens prototype developed at the University of Arizona by our team.
\label{MOIREEyeglass}}
\end{figure*}

The pioneering {\em Eyeglass} project is a very large aperture ($D$=25m--100m) diffractive space telescope concept developed by Lawrence Livermore National Laboratories \citep{Hyde1999,Hyde2002}. Eyeglass uses a transmissive, DFL as its primary DOE light-collecting element. Corrective optics in a Schupmann configuration are used to provide broadband (470~nm to 700~nm) diffraction-limited imaging at visible wavelengths \citep[][]{Bernet2017}. A 5~m diameter segmented (72 panels) prototype was built and successfully tested (see Figure~\ref{MOIREEyeglass}) and extrapolations suggest 2--3 orders of magnitude lower weight-per-aperture-area than that for HST's primary mirror (180 kg/m$^2$ -- in itself relatively low-weight due to its ``egg crate'' structure). For Eyeglass options for the primary DOE material are thin sheets of glass or silica and films of polymers such as CPI or other fluorinated polyimides.

In a newer concept, the DARPA-funded Ball Aerospace project MOIRE (Membrane Optical Imager Real-Time Exploitation) aims to develop a 20-meter aperture telescope using circular diffractive optics. MOIRE plans to deploy with a dedicated launch and capture live video and images of terrestrial targets in narrow spectral bandwidths ($\sim30$~nm, see \citealt{Hansen2013}). The MOIRE project has been under development since 2010, and results from a 5m-scale brassboard instrument have recently been reported \citep{Atcheson2014}. The published plan is to make a glass master and then replicate membranes directly from it. The process used to make the MOIRE master is a multi-level lithographic approach. 

Although Eyeglass and MOIRE demonstrate the interest and potential for very large aperture and relatively low cost space telescopes based on diffractive optics, neither of the designs are optimal for astrophysical applications, where broad wavelength coverage, faster optics (short relative focal lengths), and long operational lifetime are important.

\section{A Large-Scale Survey for Atmospheric Biosignatures}    
\label{BiosignatureSearch}

The atmospheric characterization of extrasolar planets requires separating the light emitted or transmitted by the planet's atmosphere from that of its host star, which is -- given the nearly ten orders of magnitude visual brightness difference between the Earth and the Sun -- a major observational challenge. Planet characterization methods separate planet light from its host's light spatially \citep[e.g.,][]{Siegler2018a} or utilize temporal modulations. Spatial separation (extremely high-contrast imaging) places extreme demands on the optical quality of the imaging system and will require high-performance coronagraphs \citep[][]{Trauger2007,Guyon2014} or an external occulter \citep[e.g.,][]{Cash2005,Martin2018,Seager2018} to suppress starlight by a factor of 10$^{10}$. The alternative approach, studies of temporal modulations, does not require very high optical quality but relies on high photometric precision. The rapid progress in semi-conductor technology, combined with sophisticated instrument/telescope models enabled very high-precision photometry and spectroscopy from space-based telescopes. In fact, time-resolved differential photometry  measurements have been the primary method through which Earth-sized planets have been discovered and characterized to date. 

An important opportunity for the temporal separation of the planet- and starlight is offered in situations when the exoplanet passes in front of its host star. During such {\em transit events} starlight passes through and interacts with the planet's atmosphere. Differential measurements (comparing the spectra in- and out-of-transit) can be used to constrain or determine the exoplanets' atmospheric composition (see, e.g., \citealt{SeagerDeming2010}). 

A particularly important goal for exoplanet characterization is the search for atmospheric gasses that may indicate the presence of life. Remote sensing of life is a notoriously difficult task: no single, unambiguous, easily detectable gas or signature has yet been identified (or expected to be found) that would reveal the presence of life \citep[e.g.,][]{Seager2016,Kiang2018}. Instead, combinations of inferred atmospheric absorbers may be interpreted -- in the context of the planet's global properties \citep[][]{Apai2017SAG15} -- as indication for large-scale biological processes. 

In this study we describe a general concept for a space telescope with atmospheric biosignature detection as its representative science goal and, given the exploratory nature of this study, will adopt a simplistic view on biosignature detection: we will focus on the ability to simultaneous detect several key absorbers in Earth analogs (e.g., $O_2$, $O_3$, and $H_2O$ in visible to short wavelength near-infrared light). The detection of these species is not envisioned to be the final objective, but a measurement that is representative to the observational challenges and general trends characteristic to the problem.
For more in-depth discussions of the different aspects of biosignatures we refer readers to comprehensive studies of biosignatures (\citealt[][]{Seager2016,Meadows2018,Fujii2018}).

Although transit spectroscopy does not require high image quality, it can only be applied to planets with very low inclination orbits that transit their host stars as seen from Earth. Furthermore, transit spectroscopy  requires large telescope apertures for two reasons: First, while the transit signal is a fixed fraction of the stellar flux, the signal to noise ratio (SNR) of the transit signal diminishes rapidly with the distance of the system from Earth. Second, in-transit data can only be collected during the relatively brief (typically 0.5-8~h-long) planetary transit events, which occur approximately once a (terrestrial) year for sun-like stars and more frequently for less massive (and thus intrinsically fainter) stars.
Therefore, a large-scale transiting exoplanet survey will require both an efficient way to find transiting planets and a large-enough telescope that biosignatures for any given planet can be probed efficiently during only a small number of transit events. In the following we will review considerations for finding transiting exoplanets and for characterizing their atmospheric compositions.


\subsection{Science Goals of an Atmospheric Biosignature Survey}

Our baseline science goals are to simultaneously assess the diversity of possibly Earth-like planets and to carry out a statistically meaningful search for life on these worlds. The detection of biosignatures in the atmospheres of a sample of transiting Earth-sized planets (sample size $N_{pl}$) is a necessary pre-requisite to achieving these goals. 
In the following we will discuss the desired sample size, the wavelength range of interest, the distance to which transits must be detected (as a function of sample size and composition), and explore the effective telescope diameter required to carry out the survey.

\subsection{Sample Size}
\label{S:SampleSize}

The sample size (number of planets studied, $N_{pl})$ is a fundamental parameter for an exoplanet characterization and biosignature survey. {\changed The 
ideal sample size will depend upon both the specific hypotheses to be tested and the degree of background knowledge completeness.}
Our current knowledge on the properties and composition of small (1-3 R$_{Earth}$) exoplanets is very limited (typically based only on measurements of stellar irradiation, mass, bulk density \citep[e.g.,][]{Rogers2015,Fulton2017,Grimm2018}). 
Such limited data and large uncertainties may only allow probabilistic assessment of the possible nature of the detected planets \citep[e.g.,][]{Bixel2017,Catling2018}, although this assessment can be supplemented by statistical predictions from planet formation models \citep[][]{,Apai2018_WP1}. Models predict that rocky planets with compositions different from Earth may be common and some classes may be habitable (e.g., waterworlds: \citealt[][]{Kite2018}). Our current understanding of the diversity of exoplanet formation and evolution scenarios does not provide a robust basis for predicting the diversity of present-day exoplanet population. 

{\changed \citet{Checlair2019} argue that, fundamentally, there are two approaches to exploring the diversity of exo-Earths and to search for life: one in which a direct extrapolation of solar system-type exoplanets are tested (i.e., search for Earth analogs), and one in which general hypotheses are tested. 
\citet{Checlair2019} show that the expected yield of LUVOIR could enable a statistical test of the predicted relationship between the carbon cycle efficiency and insolation, but our larger sample would allow us to observe this relationship over more dimensions such as planet size and stellar mass.
A somewhat similar argument is laid down by \cite{Ramirez2019}, who argue that understanding rocky planets as systems and identifying those that harbor life will likely require the study of hundreds, if not thousands of planets: ``We also suggest that next-generation missions are only the beginning of a much more data-filled era in the not-too-distant future, when possibly hundreds -- thousands of HZ [habitable zone] planets will yield the statistical data we need to go beyond just finding habitable zone planets to actually determining which ones are most likely to exhibit life." Also consistent with this argument is the exoplanet community report by \citet[][]{Apai2017SAG15}, that highlights the need for large samples of exoplanets to be studied in order to build up the contextual knowledge necessary for understanding abiotic diversity and outliers of potentially biological nature.}

The variety of processes and the range of key parameters involved in planet formation and subsequent evolution suggest that there is a strong likelihood for a great diversity in rocky planets -- even without considering the possible impacts of extraterrestrial life on the evolution of the planetary atmosphere and climate. With this motivation we set the target sample size of potentially habitable planets to $N_{pl}=1,000$ for our science case. This sample size should be thought of as a sample size representative to the anticipated complexity of the parameter space exo-Earths may occupy, rather than a well-determined value.

\subsection{Definition of the Planet Sample}
\label{S:SampleDefinition}

{\changed For transmission spectroscopy, the target selection fundamentally impacts the effective telescope {\changed collecting area (aperture)} required and it is, thus, a key property of any survey. In this section we explore {\changed several potential target samples and determine the distances at which the furthest targets stars in each are located}. The target selection consideration described below aims to demonstrate that viable options exist for {\changed multiple} target samples, rather than to provide {\changed a} final, optimized sample; additional considerations will lead to different target sample{\changed s} but should not affect the general feasibility.

In the following we will consider four different samples defined by host star spectral type distribution. For each of the samples we {\changed }estimate the distance of the furthest host star, its brightness, and the relative and absolute amplitudes of the transit spectroscopy signal. 
In order to explore the range of target star distances we first calculate} the expected number of transiting Earth-sized planets as a function of stellar mass ($M_*$, a proxy for stellar spectral type), distance ($d$) based on the volume density of stars ($\rho(M_*)$), the occurrence rate of Earth-sized habitable zone planets ($\eta_\earth(M_*)$), and the probability of these planets transiting ($P_{tr}$):

\begin{equation}
   N(d,M_*) = \frac{4}{3} \pi d^3 \times \rho(M_*) \times \eta_\earth(M_*) \times P_{tr}.
\label{Eq:NumTransiting}
 \end{equation}

The transit probability is $ P_{tr}=R_*/a_{HZ}$, where $a_{HZ}$ is the semi-major axis of the habitable zone. The volume density of stars of different spectral types is calculated based on the RECONS sample of the local 10~pc volume \citep[][]{Henry2018}. We follow the model of \citet{Kopparapu2013} for the habitable zone boundaries, adopt their optimistic boundaries, and assume that the planets have transit probabilities that are the mean of the transit probabilities of planets at the inner and outer boundaries of the habitable zone. With a self-developed program  -- utilizing the astropy \citep{Astropy2018,Astropy2013} and numpy \citep{Numpy2011} libraries -- we calculated the properties of stars and planets in potential target samples, including distance, transit probability, transit depths, apparent brightness, and relative and absolute transit signals.
Table~\ref{Table:SampleAssumptions} captures the key assumptions of our calculations.

\begin{figure*}[!htbp]
\epsscale{1.0 }
\plotone{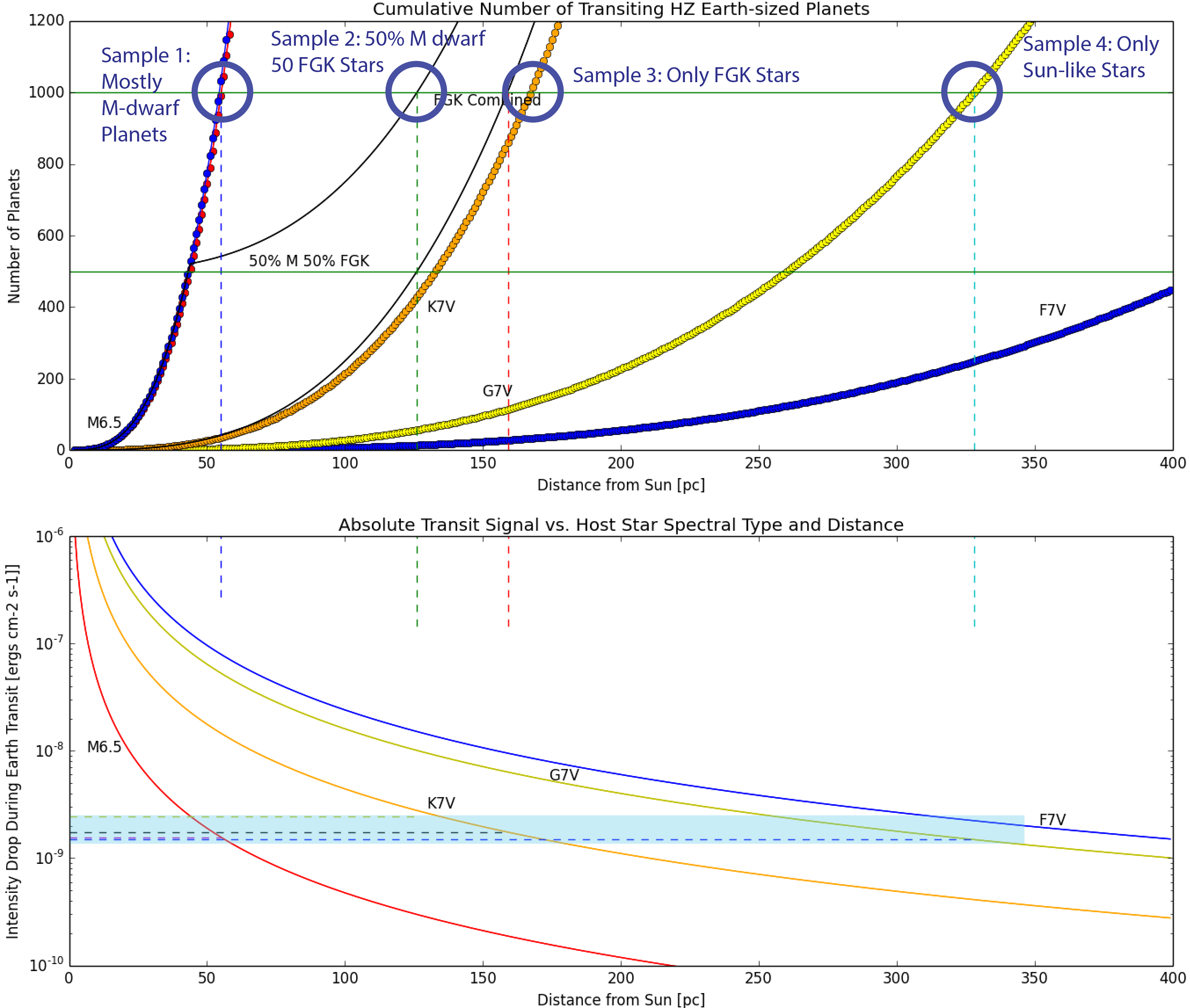}
\caption{{\em Top panel:} The cumulative number of transiting, Earth-sized planets around F stars (blue curve), G stars (yellow curve), K stars (orange curve), and M-type host stars (red curve) as a function of distance from the Sun. Four possible samples of 1,000 transiting Earth-sized planets are shown: Sample 1 includes the closest 1,000 transiting Earth-sized planets and will be dominated by planets orbiting M-dwarf host stars. Sample 2 will include up to 500 M-dwarf planets. Sample 3 includes only planets orbiting FGK-type stars. Sample 4 only includes broadly sun-like stars. The maximum distances of stars in the four samples are 55~pc, 130~pc, 160~pc, and 330~pc. {\em Lower panel:} Absolute signal (intensity drop during transit) as a function host star distance and spectral type (for transits of Earth-sizes planets). The vertical dashed lines show the distance of the most distant star in each sample. The shaded region shows the level of the faintest absolute signal in the four samples (transits around the faintest host stars). \label{SampleSize}}
\end{figure*}

In Figure~\ref{SampleSize} we show the cumulative number of transiting, Earth-sized, habitable zone planets as a function of distance and host spectral type (dotted curves). We also show four possible target samples selected from these planets. {\em Sample 1} (red dotted curve) consists of the closest 1,000 transiting Earth-sized habitable zone planets. This sample is dominated by planets orbiting M-type host stars (blue dotted curve), but the stars in this sample are confined to within $\sim$55~pc within the Solar System. Our {\em Sample 2} consists of up to 500 M-dwarf planets, supplemented by planets around FGK hosts. Stars in this sample are within $\sim$130~pc of the Solar System. Our {\em Sample 3} consists of only planets of FGK-type stars (no M-dwarf planets). In this sample, to reach 1,000 planets, we need to include stars up to 160~pc. 
In our {\em Sample 4} we included only planets orbiting broadly sun-like stars (G spectral type). With this criteria our targets are located up to 330 pc away (see~top panel of Figure~\ref{SampleSize}). Table~\ref{T:TargetSample} summarizes the spectral type distribution in each sample, as well as the distance of the furthest star and the brightness of the faintest star in each sample. The table also provides the period range of the habitable zone planets for the target sample.

Exoplanet transit observations are relative measurements: the signal strength measured is relative to the apparent brightness of the host star and the ratio of the planet's and the star's projected surface areas. As the four samples defined in our study contain stars of very different sizes, brightness, and typical distances, the comparisons of the samples in terms of the ease of detectability is non-trivial. The bottom panel of Figure~\ref{SampleSize} shows the absolute flux density difference during the transit of an Earth-sized planet around the target stars, as a function of spectral type and distance. We note that the flux density difference shown on the $y$ axis is continuum transit depth and not specific spectral features (which are typically several orders of magnitude fainter). For each of the curves (for stars of different spectral types) we also mark the faintest (most distant) stars in the target samples (as determined and marked in the top panel of the same figure). The dashed lines parallel with the $x$ axis denote the flux density levels corresponding to the transits around the faintest stars in the sample.

Interestingly, the out-of to in-transit difference signal (in flux density) predicted for the faintest stars in all four samples falls within a relatively narrow range: 2--5$\times10^{-9}~erg\,cm^{-2}\,s^{-1}$. This finding suggests that the absolute differential photometric precision required to detect a {\em single} transit of an Earth-sized planet around the {\em faintest} host stars in each sample are very similar between the samples; and that the precision floor is primarily set by the target sample size. However, because the number of transits of habitable zone planets that occur within a given time window, as well as the brightness of the host stars (and therefore the required  precision) are strongly spectral type-dependent, the four samples discusses above are not equally well suited for a survey. We will consider these factors further in Sections~\ref{S:SpecSim}.

\begin{deluxetable}{cccccccc}
  \caption{Properties of target samples. The target samples include very close to 1,000 habitable zone earth-sized planets, but reflect different choices in the exoplanet host stars. \label{T:TargetSample}}
\tablehead{\colhead{Sample}&\multicolumn{4}{l}{Planets around Stars}&\colhead{Max. Dist.} &  \colhead{Faintest}& \colhead{HZ Peri-} \\
\colhead{ID}&\colhead{M}&\colhead{K}&\colhead{G} & \colhead{F}  & \colhead{[pc]} & \colhead{I-mag}& \colhead{ od [day]}} 
\startdata 
1 & 1000 & 0  & 0 & 0 &  55&  16.5 & 4--18 \\
2 & 500 & 438 & 58& 14&126 &  15.9 & 4--1370 \\
3 & 0 &876    &116& 29& 159&  12.6 & 60--1370 \\
4 & 0 & 0 & 1000 & 0 & 328 &  12.0 & 200--740 \\
\enddata
\end{deluxetable}

Considering the connections between stellar luminosity, habitable planet occurrence rate, transit probability, transit depth, and the volume (number) density of stars of different spectral types, we evaluated target definition choices and the resulting absolute signal strength. Our three key conclusions from the target selection study are: (1) In the case of a single transit the resulting absolute signal strength -- corresponding to the transit of an earth-sized planet -- is insensitive to the spectral type distribution of the target stars. (2) For a survey limited by single-transit absolute signal strength, the key parameter of the survey definition is the sample size. (3) Considering the order-of-magnitude shorter orbital periods of habitable zone planets around M-type host stars and the deeper (relative) transit depths, for a survey limited by total telescope time and by relative photometric precision, a sample rich in M-type host stars may be advantageous. Such a sample is also optimal if the transiting planets must be located by the survey before the spectroscopic characterization can began (due to the shorter transit periods around M-dwarfs, much shorter temporal coverage is required to detect or exclude transits). 
We note, however, that more detailed future studies are warranted to assess the impact of other factors not considered here, including stellar activity and the photometric variability it causes, and low-level stellar contamination of the transit spectra due to stellar heterogeneities \citep[][]{Rackham2018}, which affect sun-like stars much less than M-dwarfs \citep[][]{Rackham2019}.

\subsection{Transmission Spectrum Simulations}
\label{S:SpecSim}

We utilize the Planetary Spectrum Generator\footnote{\url{https://psg.gsfc.nasa.gov}} \citep[PSG,][]{Villanueva2018} to provide an approximate assessment of the expected quality of the spectra obtained by the telescope concept and explore what equivalent telescope diameter is required for the Nautilus Observatory to ensure the detectability of $O_2$, $O_3$, $H_2O$, and $CO_2$ absorption features in our target sample. The PSG is capable of simulating observations of model atmospheres from a variety of viewing geometries, either for planets within the Solar System or around distant host stars of varying spectral types.

We run the PSG for the four models presented in Table \ref{T:PSG}. These include a mid-M dwarf and solar analog, each at a nearby and far distance based on the relative abundance of transiting planets in Figure \ref{SampleSize} {\changed (10~pc and 50~pc for the M dwarf, 100~pc and 300~pc for the G dwarf, see Table~\ref{T:PSG})}. In all models, the planet is an Earth-sized, Earth-mass planet with the PSG's default profile for the Earth's atmosphere. The orbit is placed in the middle of the habitable zone as estimated from the host star's properties by \citet{Kopparapu2014}. The viewing geometry is set to `Observatory' mode, with a planetary inclination of 90$^\circ$ and orbital phase of 180$^\circ$, and the viewing distance set to one of the four values in Table \ref{T:PSG}.

The PSG can incorporate both Poisson and instrumental (e.g, readout) sources of noise, but we ignore the instrumental noise to focus on the effect of aperture size. The Poisson noise is calculated for a telescope array, each taking 1s exposures. We set the number of exposures such that the total exposure time equals the transit duration, and generate a spectrum from 200$-$1,800~nm with 1~nm resolution for each of the four models in Table \ref{T:PSG}. 

The output of the PSG for the viewing geometry described above is the fraction of light blocked by the planet as a function of wavelength, with uncertainty estimates in each bin. The uncertainty estimates correspond to the amount of light collected over the entire transit duration. We take this to be the uncertainty on the transit depth, but we inflate the uncertainties by 20\% to account for limb darkening degeneracy. We assume that each planet will be observed in transit multiple times -- 10 times over the course of a year for the M dwarf planets, and 5 times over the course of $\sim 7$ years for the G dwarf planets {\changed -- and we reduce the uncertainties by $\sqrt{N_\text{obs}}$ for $N_\text{obs}$ observations. The justification for the number of revisits for each type of host star is explored in Section \ref{sec:revisits}.} We varied the telescope array configuration to explore the equivalent collecting area that satisfies our science goal.

\begin{figure*}[!htbp]
\epsscale{1.15}
\centering
\includegraphics[width=0.49\textwidth]{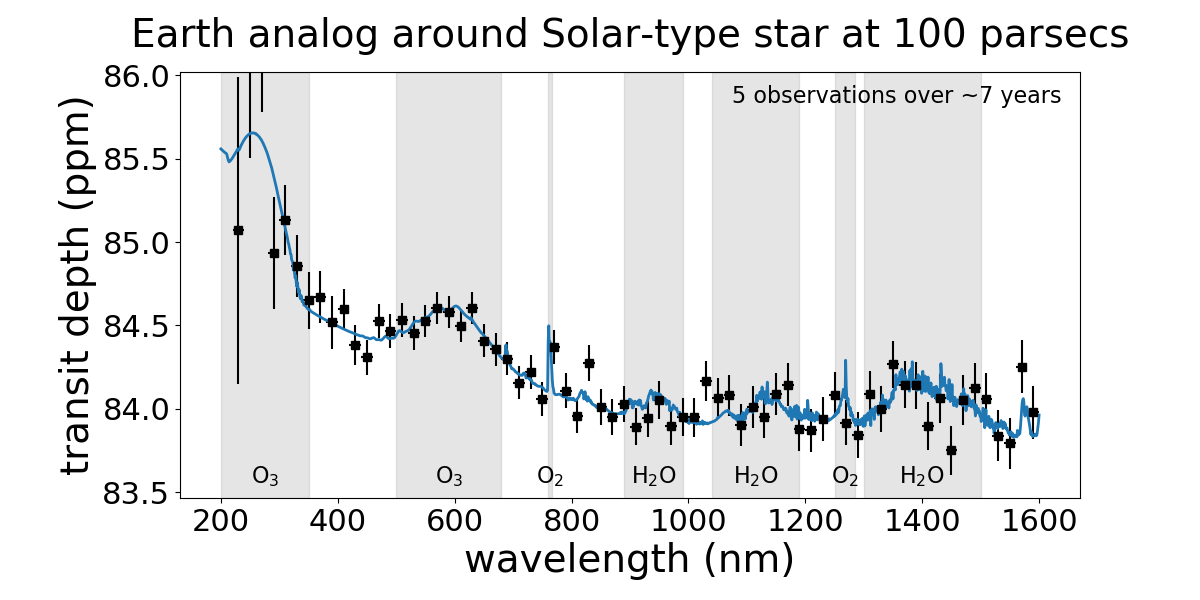}
\includegraphics[width=0.49\textwidth]{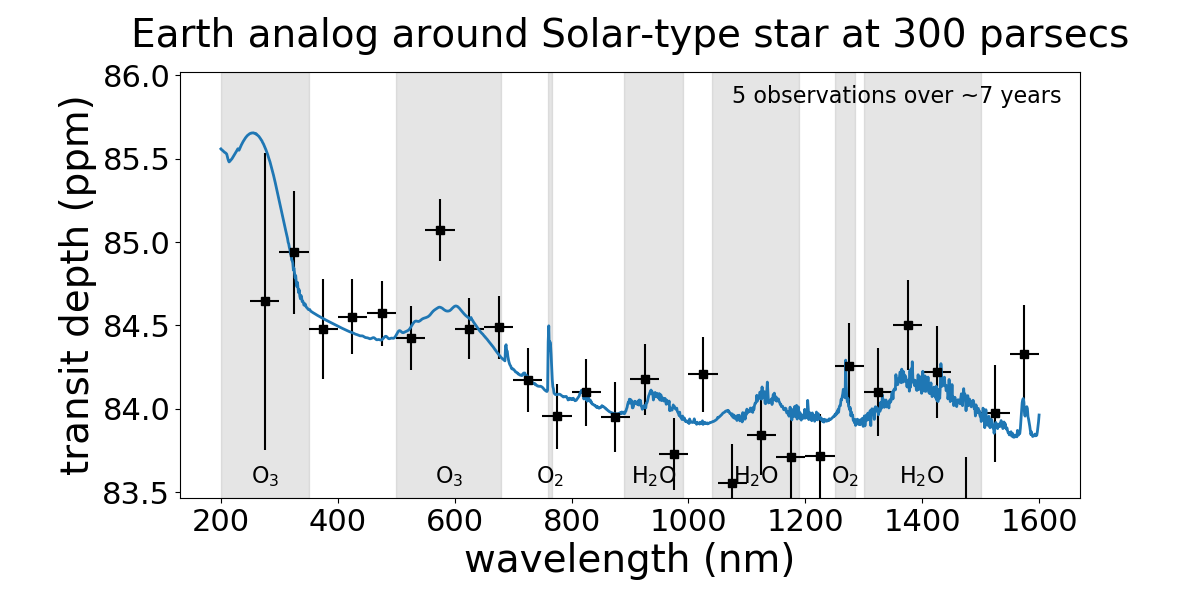}
\includegraphics[width=0.49\textwidth]{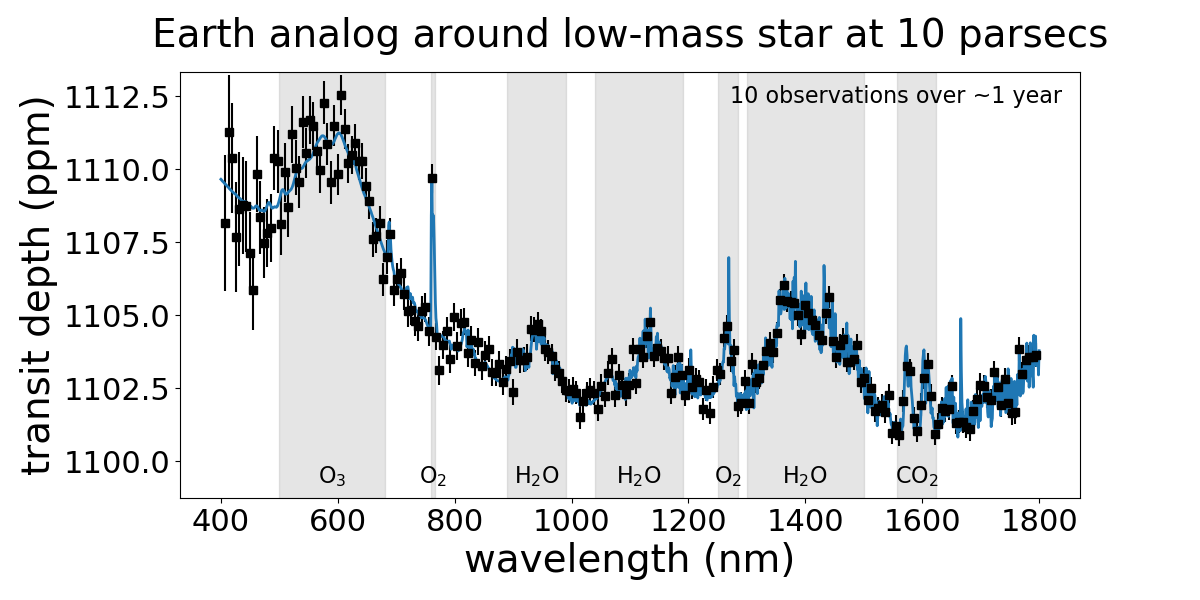}
\includegraphics[width=0.49\textwidth]{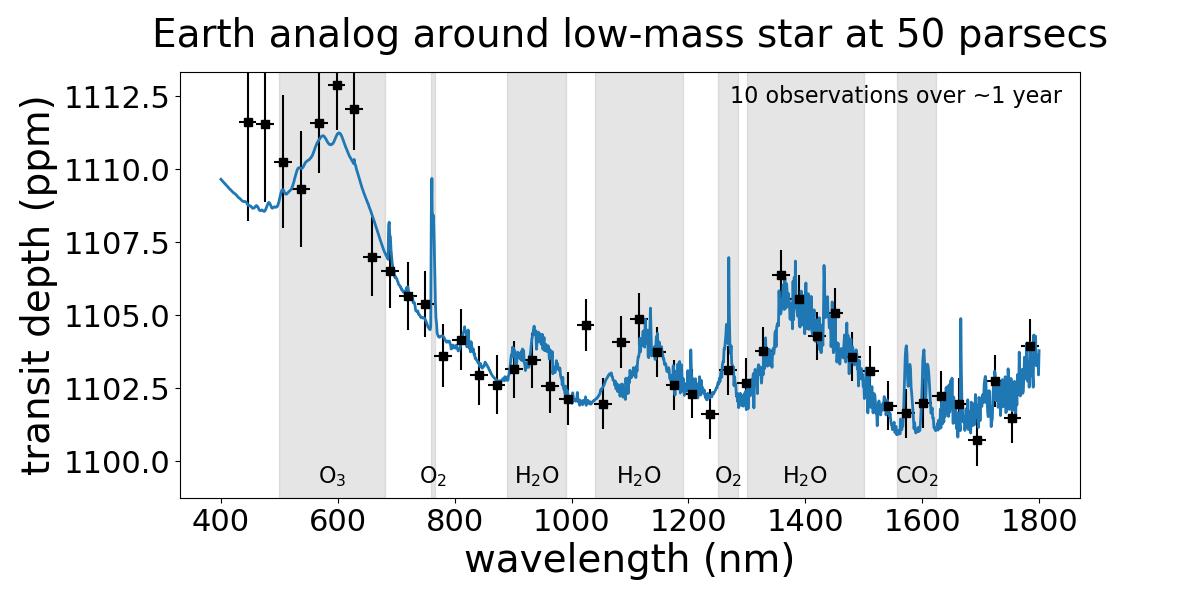}
\caption{Simulated Nautilus Observatory spectra of nearby and far targets (Earth-sized habitable zone planets) around low-mass stars (M-dwarfs) and sun-like stars (G dwarfs), assuming 35 telescopes with 8.5m-diameter apertures. The light-collecting power equivalent to that of a $D\sim$50m telescope enables the detection of H$_2$O and O$_3$ in $\sim$1,000 Earth analogs. The same configuration often also enables detection of O$-2$ and, in hundreds of simulated Earth-analogs around nearby low-mass stars, CO$_2$ absorption. \label{SimSpectra}}
\end{figure*}

Finally, in each spectral bin we draw a value from the normal distribution defined by the uncertainty to simulate the transmission spectrum achieved for each of the four models. The simulated spectra are presented in Figure \ref{SimSpectra}, and binned for visibility according to Table \ref{T:PSG}. We find that with a 35-element array of 8.5m-diameter telescopes three partial biosignatures -- $O_3$, $H_2O$, and potentially $O_2$ -- could be identified in the atmosphere of an Earth twin orbiting a nearby solar-type star. For nearby M-type hosts, all of these plus $CO_2$ could be identified with high confidence. Furthermore, detection of $O_3$ and $H_2O$ could be achieved for the most distant planets in the sample, allowing for a statistical analysis of the presence of biosignatures for hundreds of planets. Although our simulations show a broader wavelength range, we conclude that a narrower range (such as 500--1,000~nm) is sufficient for simultaneously detecting three key atmospheric components ($O_3$, $H_2O$, $O_2$) of an Earth analog.

\begin{deluxetable*}{cccc|ccccc}
  \caption{Parameters of the four models run through the Planetary Spectrum Generator. In all cases we use the default profile of an Earth-like atmosphere with $R_{pl} = 1 \, R_\oplus$. The uncertainties shown are calculated for a 35$\times$8.5\,m array of telescopes with continuous 1s exposures over the duration of the transit, and we inflate the uncertainties by 20\% to account for limb darkening degeneracy. Finally, we reduce the uncertainties to account for the benefit of multiple transit observations. \label{T:PSG}}
  \tablehead{\colhead{Host type} & \colhead{$M_*$} & \colhead{$R_*$} & \colhead{$T_\text{eff}$} & \colhead{Distance} & \colhead{Semi-major axis} & \colhead{Transit duration} & \colhead{\# of observations} & \colhead{Binning}\\\colhead{} & \colhead{($M_\odot$)} & \colhead{($R_\odot$)} & \colhead{(K)} & \colhead{(pc)} & \colhead{(AU)} & \colhead{(min)} & \colhead{} & \colhead{(nm)}}
  \startdata 
    M  & 0.20 & 0.28 & 3020 & 10 & 0.12 & 170 & 10 & 6 \\ \hline
    M  & 0.20 & 0.28 & 3020 & 50 & 0.12 & 170 & 10 & 20  \\ \hline
    G  & 1.00 & 1.00 & 5780 & 100 & 1.3 & 900 & 5 & 20  \\ \hline 
    G  & 1.00 & 1.00 & 5780 & 300 & 1.3 & 900 & 5 & 50  \\ \hline
  \enddata
\end{deluxetable*}

\subsection{Number of Visits} \label{sec:revisits}
{\changed Transit spectroscopy benefits from repeated transit observations{\changed. C}ombining the transmission spectra from several transits decreases uncertainties by $\sim\sqrt{N_\text{obs}}$. This strategy has been employed several times from the ground \citep[e.g.,][]{Rackham2018,Bixel2019} and space \citep[e.g.,][]{Kreidberg2014,deWit2018,Zhang2018}{\changed. A} similar approach could be followed to achieve low-to-moderate signal-to-noise spectra of the atmospheres of Earth-sized planets in systems such as TRAPPIST-1 with JWST \citep{Batalha2018}.

When possible, it is therefore optimal to observe a planet's spectrum during every transit. A candidate exo-Earth in the middle of the habitable zone of a Sun-like star will transit once every $\sim 1.5$ years, and the duration of each transit observation (including baseline measurements) would be up to $\sim 15$ hours. Observing all of the planets with G-type hosts in Sample 3 during every transit would cost $\sim 1600$ hours ($\sim 2$ months) per year; this commitment {\changed of time} would be justified by the value of characterizing $> 100$ ``true'' Earth twins.

However, planets orbiting low-mass stars are far more common with orbital periods of $< 30$ days. The required amount of time for a transit observation is still significant (up to $\sim 5$ hours), and to observe every transit of, for example, 500 such planets (Sample 2) would require 5-10$\times$ more {\changed } observing time per year than {\changed is} available. Fortunately, the short orbital periods and larger transit depths of such planets will allow an observer to achieve a high signal-to-noise transmission spectrum with $\sim 10$ observations in less than a year, and the full sample of planets with M-type hosts could be studied sequentially over a decade.

Finally, since the signal-to-noise scales with $\sqrt{N_\text{obs}}$, there are diminishing returns when combining {\changed large number (tens) of} transit observations for a given target. Nevertheless, closer to $\sim 100$ visits could be scheduled for a handful of nearby interesting low-mass systems (e.g., TRAPPIST-1, \citealt{Gillon2017}) to enable very high signal-to-noise spectroscopy, but we assume that the amount of time spent on such systems will be negligible in the overall scope of the mission. We use these arguments to set the number of combined transit observations for the results shown in Figure \ref{SimSpectra}, and we see that in about eight years we can achieve {\changed statistically significant positive} results for the samples suggested in Table \ref{T:TargetSample}.}

\section{Preliminary Science Requirements and Operations}
\label{SciReq}
For the purposes of this exploratory study we adopt the following high-level, preliminary science requirements as: 1) An effective light-collecting aperture equivalent to a $D=50m$ telescope. 2) A wavelength coverage that includes key biosignature absorption bands, such as 450-1,000~nm. 3) near photon-noise limited telescope performance (including temporal stability) after post-processing; and 4) low-resolution spectroscopic capability with spectral resolution between 6 to 20 nm (corresponding to $R=\lambda / \Delta \lambda=50-170$).

In Table~\ref{T:Requirements} we summarize the science requirements of a statistically meaningful atmospheric biosignature survey. In the following section (\S\,\ref{S:NautilusArray}) we will describe our novel concept for a space telescope array capable of carrying out the biosignature survey described here. We note that these requirements are intended to be representative and not conclusive; future, more comprehensive studies will be required to fully define the requirements for an actual flight mission.

\begin{deluxetable*}{lccc}
  \caption{Baseline science requirements for the Nautilus Biosignature Survey. \label{T:Requirements}}
  \tablehead{\colhead{Parameter}&\colhead{Value/Range}&\colhead{Science Driver}&\colhead{Ref.}} 
  \startdata 
Num. Exo-Earth Candidates  & 1,000 & Expected exoplanet diversity, statistically meaningful results & \S \ref{S:SampleSize}      \\ 
Faintest host stars probed & I=16.5 & Furthest/coolest star in sample & \S \ref{S:SampleDefinition}      \\ 
Wavelength Range       & 0.45--1.0~$\mu$m & H$_2$O, O$_3$ molecular bands  & \S \ref{S:SpecSim}      \\ 
Photometric Precision & $\sim$1~ppm & Absorption feature depth & \S \ref{S:SpecSim}      \\ 
Spectral Resolving Power & $\lambda / \Delta \lambda$=50-170 & Molecular band widths & \S \ref{S:SpecSim} \\
\enddata
\end{deluxetable*}
  
\subsection{Nautilus Telescope Array}
\label{S:NautilusArray}
In this section we introduce the Nautilus Observatory, a novel telescope array concept developed to meet the science requirements (see \S~\ref{SciReq} and Table~\ref{T:Requirements}) of the large-scale atmospheric biosignature survey we described in \S~\ref{BiosignatureSearch}. {\changed The Nautilus concept described here is not a complete, final, and fully optimized mission design.}  {\changed Rather, it is a notional design that highlights the potential of  novel large-scale diffractive optics to answer key astrophysical question. In this manuscript we focus on technical opportunities and challenges unique to MODE lens-based space observatories or the Nautilus concept. We do not address technical aspects that are shared with other space observatories.}
We will first review the baseline concept for the observatory and its operations, review individual unit telescope architecture, their launch and deployment, and various fundamental considerations.

\subsection{Nautilus Array Baseline Concept}
{\changed Our study described in \S\,\ref{S:SpecSim} established that a telescope system with a light-collecting power equivalent to a single 50~m diameter aperture will be required for the biosignature survey. As no such large diameter single telescope is realistic to launch in the foreseeable future, our  mission design envisions {\changed multiple} unit telescopes that combine light {\em non-coherently} to match the light-collecting power of {\changed a} 50m telescope. Our notional Nautilus concept utilizes an array of ultra lightweight, {\changed very-large-aperture}, and low-cost unit space telescopes with powerful light-gathering capabilities. 
In {\changed considering} the diameter of the individual unit telescopes{\changed,} we {\changed adopted a size} that is consistent with the largest rigid (non-folding) diameter that can be launched in the next decades: specifically, we {\changed adopted} a diameter of D=8.5m, about 5\% smaller than the maximum {\changed} inner dynamic {\changed envelope} diameter of the fairings of the next-generation {\changed vehicles} (SpaceX/BFR, NASA SLS B2). In order to match the light-collecting power of a 50m telescope, 35 such unit telescopes will be required.
We note that the design presented here is not sensitive to the specific diameter of the unit telescope's diameter: if unit telescopes with smaller apertures are used, the number of units telescopes can be increased to keep the total area constant. Future trade studies will be required to verify that the 35$\times$8.5m configuration is an optimal choice.}

This novel telescope architecture is {\changed potentially enabled} by the rapid progress in replicated multi-order diffractive engineered material (MODE) lenses which have the potential to replace primary mirrors.
Their incoherently combined light (digitally co-added signal) collecting capability will equal that of a single 50~m mirror diameter space telescope.

{\changed We envision Nautilus to operate primarily in follow-up transit spectroscopy mode, but also to have the capability for exoplanet transit searches.} The combined operations will allow discovering and characterizing a very large number of habitable zone earth-sized transiting exoplanets.

{\em Transit search mode:} {\changed The Nautilus Observatory will benefit from multiple powerful transiting exoplanet search missions that precede it (e.g., Kepler, \citealt{Borucki2010}; TESS, \citealt{Ricker2015}; PLATO \citealt{Rauer2014}, which are expected to identify tens of thousands of transiting exoplanets. Nevertheless, the Nautilus Observatory will be capable of searching for transits on its own, enlarging the potential target sample.} Operating independently of each other, unit telescopes will monitor potential exoplanet host stars in the target sample, and through their parallel operation {\changed they will have the potential to} carry out the most sensitive and most comprehensive {\changed transiting} exoplanet search yet. The unit telescopes will use their smaller (2.5m diameter) MODE lens -- optimized for wide field of view imaging -- for the transit search. The transit search component will greatly expand the number of known transiting habitable zone earth-sized planets. 

{\em Follow-up transit spectroscopy mode:} During known transit events all unit telescopes will record the transmission spectrum of the same planet using their larger, 8.5m diameter MODE lenses. The signal measured by the individual unit telescopes will be combined non-coherently (by digitally co-adding), enabling the confident detection of major atmospheric absorbers (O$_2$, O$_3$, H$_2$O) in Earth twins up to about 300~pc. 

{\changed The non-coherent combination of signal, as planned in the Nautilus Observatory concept, does not require formation flying for the unit telescopes. As light is combined non-coherently, the relative locations of the individual telescopes during the observations (as long as the target star is visible) are not important.}

\begin{figure*}[!htbp]
\epsscale{1.15}
\plotone{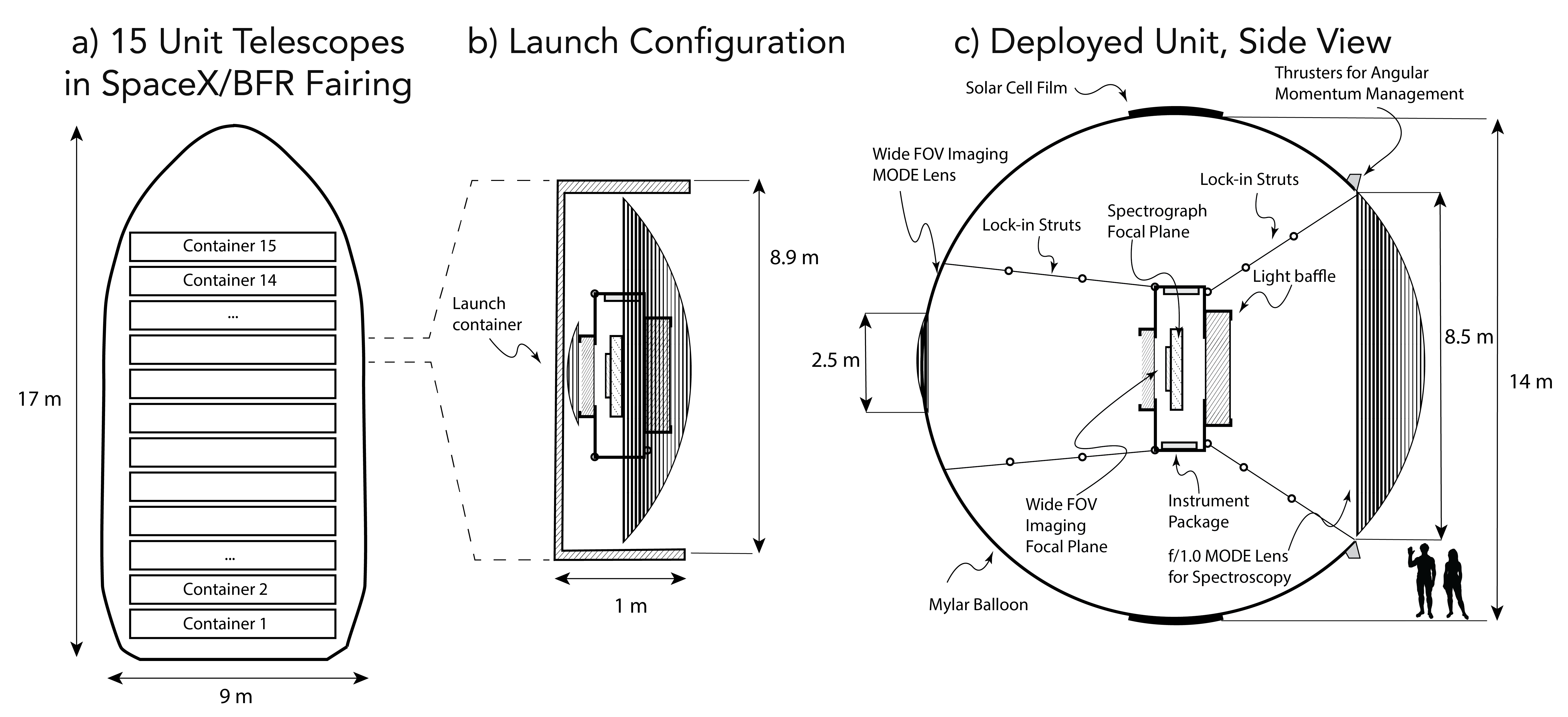}
\caption{Baseline concept: a) The compact launch configuration allows 15 Nautilus units (each 8.5m lens diameter) to be launched in a single launch vehicle (here SpaceX BFR). b) Single unit in launch configuration. c) Once in orbit a gas canister inflates a mylar balloon, deploying the instrument package and the MODE lens. Lock-in struts provide additional mechanical stability and longevity. A second, smaller MODE lens provides parallel imaging capabilities with large field-of-view, ideal for exoplanet transit search or deep imaging surveys. 
\label{ThreeStages}}
\end{figure*}

\begin{figure*}[!htbp]
\epsscale{1.25}
\plotone{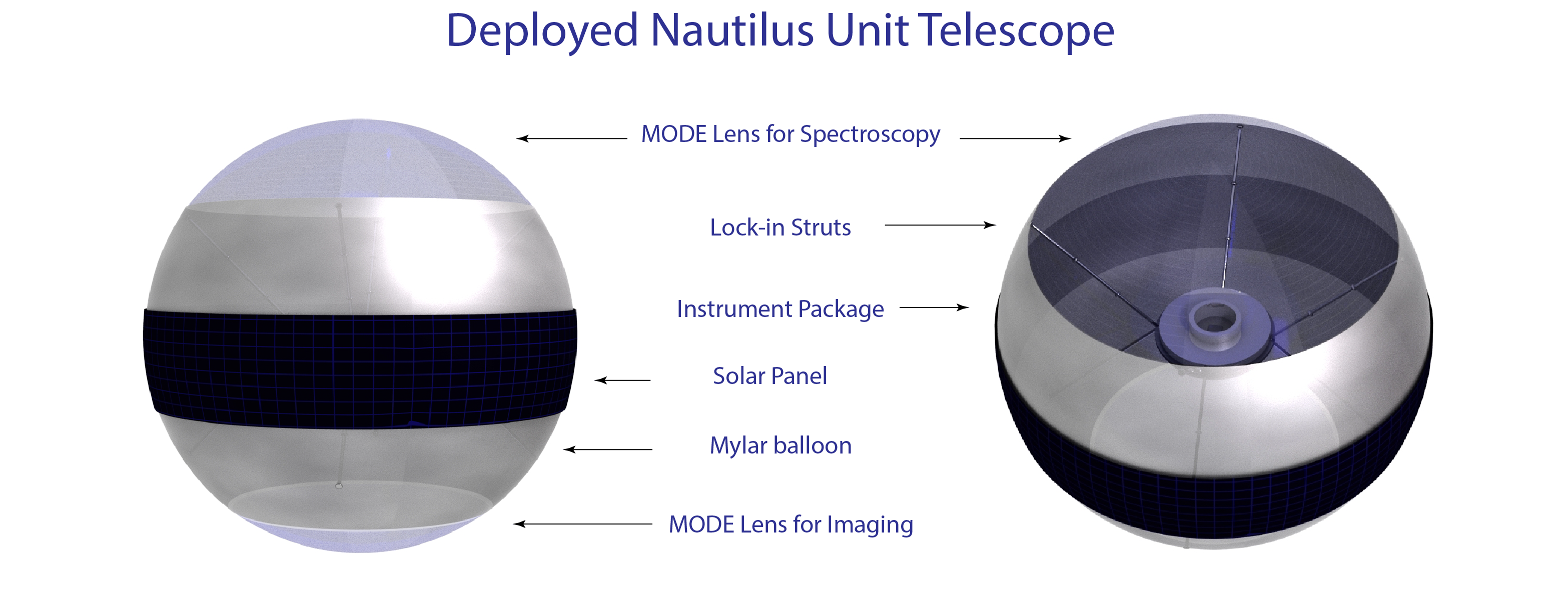}
\caption{A single 8.5m lens-diameter Nautilus unit telescope after deployment. The instrument payload is visible at the geometric center of the unit; a light baffle controls off-axis and internally scattered/reflected light. Lock-in struts provide mechanical stability. The solar cell film is visible as the equatorial dark belt.  \label{3DModel-Telescope}}
\end{figure*}

\begin{figure*}[!htbp]
\epsscale{1.25}
\plotone{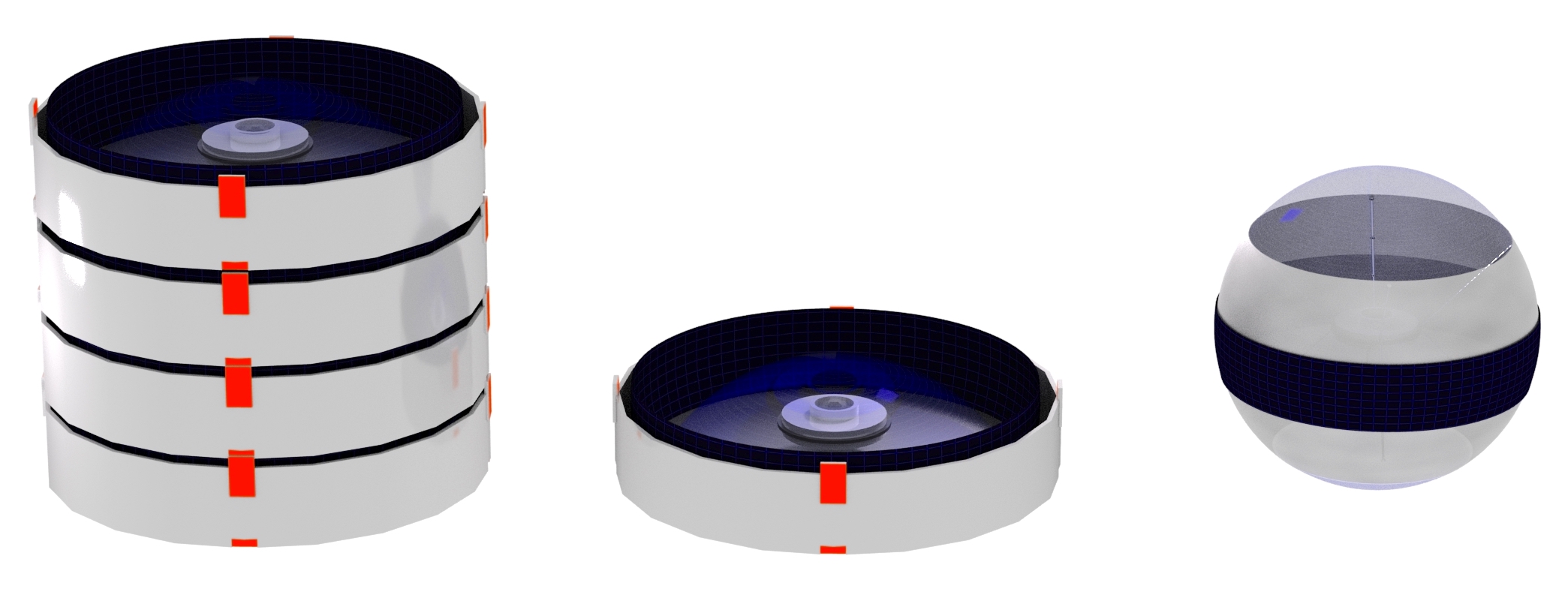}
\caption{Rendering of the Nautilus launch containers ({\em left}), a single unit telescope in the launch container ({\em middle panel}), and a single deployed unit ({\em right}).  \label{3DModel-Containers}}
\end{figure*}

\begin{figure*}[!htbp]
\epsscale{1.15}
\plotone{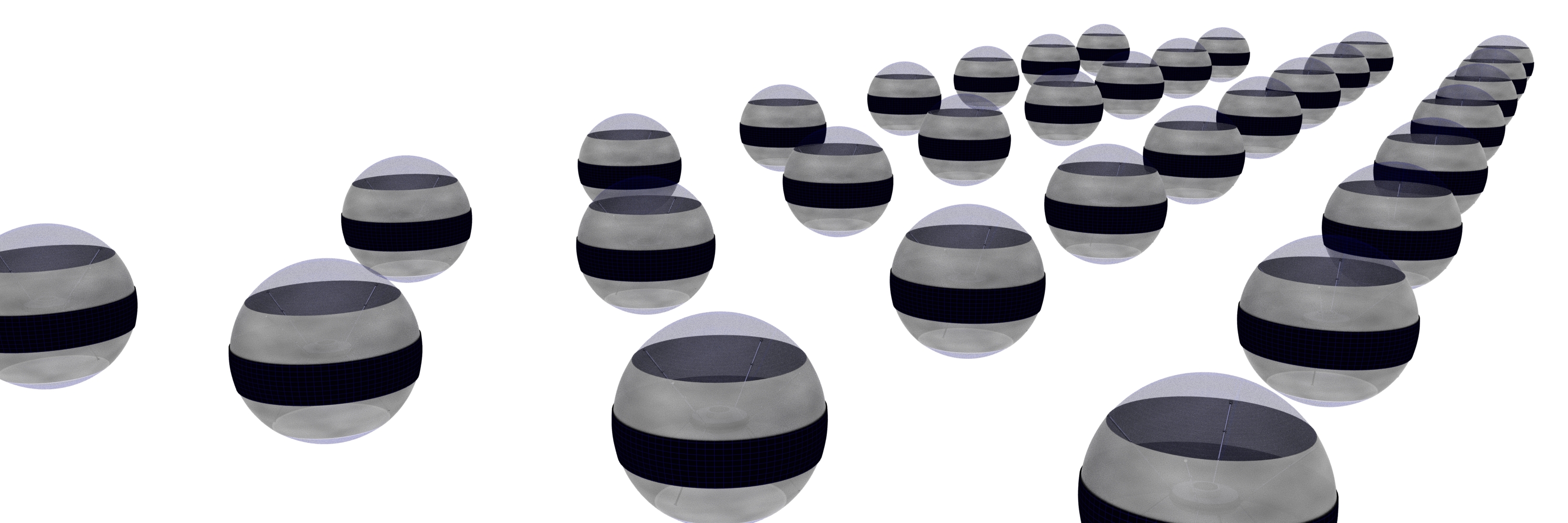}
\caption{The Nautilus Array will consist of about 35 unit telescopes, providing a combined light collecting area equivalent to a single 50~m diameter telescope.  The units do not need to fly in formation or even be located close to each other, as the intensity measurements are combined non-coherently (i.e., co-added). \label{3DModel-Array}}
\end{figure*}

\subsection{The Nautilus Unit Telescope}

Figures~\ref{ThreeStages},~\ref{3DModel-Telescope}, and ~\ref{3DModel-Containers} illustrate our baseline concept for the Nautilus unit telescopes in compact launch and deployed configurations. Figure~\ref{3DModel-Array} illustrates the observatory (telescope array) in operation. Each unit telescope will use an 8.5~m diameter f/1.0 focal ratio MODE lens as the light-collecting element for the exoplanet transit spectroscopy observations; and a smaller, 2.5\,m diameter, wide field-of-view MODE lens for the photometric exoplanet transit search operations. The two lenses will focus the light on two simple instruments. Each Nautilus unit will be a stand-alone telescope equipped with two visual/near-infrared detectors and a low-resolution spectrograph optimized for the 0.45-1.0 $\mu$m wavelength range.

{\changed In this study we will not address the details of {\changed posited} instruments and detectors, as our focus is on the novel use of MODE {\changed lend} technology and the measurements strategy it enables. We note, however, that the measurements proposed {\changed herein} are generally compatible with low-noise detectors and {\changed low-spectral-resolution} spectrographs that are {\changed currently available.} If {\changed a} sufficient amount of light is collected, currently available measurement {\changed technology offer viable solutions} for the observations. For example, Kepler's CCD cameras were able to detect ppm-level modulations (after de-trending, in white light) – the same level of precision required by the observations proposed here.  We also note, that lower light throughput and detector noise can be compensated by increasing the number of unit telescopes, i.e., will not have major impact on the overall concept.}

\subsection{Orbit}

In this section we demonstrate that several satisfactory options exist for the orbit of the Nautilus Observatory. The determination of the ideal orbit will be based on a detailed assessment of the mission concept; here we provide only a preliminary discussion. The primary considerations for the Nautilus Array's orbit are low energy ($\Delta$v) access, orbital stability, very stable thermal and radiation pressure environment, and quasi-constant illumination (from Sun, Earth, and Moon).

With these considerations in mind we identified the Earth-Sun L2 point as one of the possible locations for the Nautilus Array. Stationed at the L2 point, the Sun, Earth, and Moon will be seen by the Nautilus units from nearly identical directions, allowing for a very stable radiation environment and constant communication windows.
A possible alternative, easier-to-access orbit would be a sun-synchronous, high-inclination, terminator-following low earth orbit. This orbit would provide somewhat less stable illumination and thermal environment, more limited sky coverage, but would be easier to access and to communicate with. In addition, a low earth orbit would enable passive angular momentum management (magnetic torquing rods).  The {\changed ultimate} choice of orbit will impact the spacecraft design.  

\subsection{Launch and Deployment}
\label{S:Deployment}

The Nautilus unit telescopes utilize inflatable spacecraft components for deployment, allowing the instrument package and MODE lenses to form very compact packages (see Figures \ref{ThreeStages}, \ref{3DModel-Telescope} and \ref{3DModel-Containers}).  
The unit telescopes are launched in a compact configuration, in cylindrical containers (approx. 9\,m diameter and 1\,m tall), with multiple containers positioned in a single fairing (Figure~\ref{3DModel-Containers}). 
Over the next two decades multiple launch options will exist that are suitable for the Nautilus Observatory. 
Next-generation rockets and their largest fairing will allow over a dozen units to be launched in a single payload (Figure~\ref{ThreeStages}): NASA's Space Launch System's Block 2B fairing is expected to offer a $\sim$9.1~m diameter payload envelope with a height approaching 30~m; such a fairing may be capable of launching $\simeq$24--28 Nautilus units in a {\em single} launch. SpaceX's upcoming Big Falcon Rocket (BFR) will offer a fairing with a diameter very similar to that of the SLS Block 2B long fairing, but with a probably shorter height (17 m). Correspondingly, SpaceX/BFR may be capable of launching 15 Nautilus units. In our reference design we adopt a BFR-style fairing (see Figure~\ref{ThreeStages}).

We note, that currently existing fairings are well-suited for launching smaller -- but still very capable -- pathfinder units.  For example, the currently operational SpaceX Falcon 9 accommodates a dynamic payload envelop (cylindrical) with a 4.6~m diameter and 6.7~m height. This could accommodate up to 6 Nautilus launch modules with up to $\sim4.2m$ MODE lens diameter. The 5~m diameter 'Long' fairing offered for the Atlas~V has a fairing accommodating 4.6~m diameter with a 12.2~m height, which could be sufficient for ten Nautilus units based on MODE lenses with diameters up to $\sim4.4m$. 

After orbital insertion the individual Nautilus unit telescopes separate from the fairing and from each other and begin inflation. Mechanically, the telescope deployment is driven by the inflation of a spherical mylar balloon (diameter $\sim$14~m), which will shift the two MODE lenses (in front and behind of the instrument package) forward and backward by about 6.5~m. In addition to the simple deployment mechanism, the mylar balloon will also provide stray-light control, sunshield functionality, and solar energy to the telescope. The inflation itself is a non-reversible operation, initiated by the release of a low amount of chemically inert, but relatively high atomic weight noble gas ($Kr$). Once the target shape is reached, mechanical struts lock in, fixing the telescope structure, providing long-term mechanical stability. 
The instrument packages will be aligned precisely to the focal plane of the deployed MODE lenses through observations of a reference star field.

\subsection{Power source and management}

The estimated power requirements of the Nautilus Array Unit telescopes are smaller than the power requirements of large space telescopes and, therefore, are not expected to pose significant challenge. We anticipate that the major systems requiring energy will be communications, spacecraft attitude control, and the instrument package; similar components exists on the Hubble Space Telescope, James Webb Space Telescope, and Herschel Space Observatory, all powered via solar cell arrays. 

As a baseline we estimate the operational power requirements of an Nautilus Unit telescope by scaling the power use of HST (2.8~kW) and JWST (2.2~kW). We anticipate that the power consumption of a single Nautilus unit telescope will be lower than HST and JWST due to the Nautilus units' simpler design, lack of cool instruments required for long-wavelength observations ($>1.5~\mu$m), its smaller number of instruments/subsystems, and considering more efficient electronics. 
As a reference value for power requirements we assume 1.5~kW for each telescope.

Unlike HST and JWST, Nautilus units will utilize flexible solar cell film,
which have space heritage (Venus Express) and -- due to the flexibility and low areal density -- will provide ideal structural match for the inflatable spacecraft. Nautilus units will integrate solar cell film into the inflatable balloon in a rotationally symmetrical configuration. 
Assuming that the average fraction of the solar panels' surface illuminated is $\eta_{ill}=0.3$, a solar cell efficiency of $\eta_{eff}=0.1$, an average distance of d=1~au from the Sun, and a solar constant (at 1~au) of $c_{sun}=1.37~kW/m^2$, the total surface area of Nautilus units covered in solar cell film will be $A_{sc}={1.5 kW \over c_{sun} \eta_{eff} \eta_{ill}}=36.5~m^2$ to provide 1.5 kw average power. The power will be stored in batteries, providing a stable power source also when the solar array is not illuminated. The 36.5~$m^2$ solar cell film corresponds to only about 6\% of the spacecraft's (balloon's) surface (for a balloon radius of 7~m). An orbit with greater average distance from the Sun will require somewhat larger fraction of the balloon to be covered with solar cell film. In short, available, low-weight and low-cost power source with space heritage exists that can, by a large margin, cover the energy needs of a unit telescope.

\subsection{Unit Telescope Instrument Package Volume}

Given the thin MODE lenses we estimate the available instrument package volume as a cylinder z=$0.6$~m high and with a radius of up to $r=4.0$~m, which translates into a volume of up to $\sim$30~m$^3$. In comparison, the Hubble Space Telescope's aft shroud is approximately 60~m$^3$ (radius of 2.2~m and height of 3.55~m). Therefore, a Nautilus unit telescope's instrument package volume would be overall comparable to instrument packages of existing major observatories. Given the goal to provide simple, compact, and identical instrumentation for each unit telescope, the volume available in the units is not expected to pose particular challenges. 

\subsection{Angular momentum management}

Nautilus units will use four reaction wheels (all offset for the inertial axes) to manage the rotation (pointing, tracking) of the spacecraft along three axes (a fourth wheel provides redundancy). The reaction wheels will spin in the direction opposite to the intended rotation of the spacecraft. Nautilus units -- like any spacecraft -- will be subjected to net torques (primarily due to asymmetric exposure to solar irradiation and solar wind pressure). Although the reaction wheels will ensure stable pointing during operations, Nautilus units will periodically need to dump angular momentum. We envision two possible pathways for this: a passive and an active mechanism. In the passive angular momentum management mode -- as the distribution of transiting planets is closely isotropic on the sky -- each unit telescope's observing schedule can be planned in such a way to average out torques. In the active angular momentum management mode ambient-temperature, pressurized, nitrogen is released through thrusters affixed to the exterior of the mylar balloon (possibly at the connecting points of the lock-in-struts, see~Figure~\ref{ThreeStages}). Nitrogen does not affect the planned observations and will not react with or freeze onto the spacecraft.
Given the unusually symmetric architecture of the unit telescopes, net torques will be lower than they are for most other spacecraft architectures, resulting in a much lower than typical rate of angular momentum accumulation.

\subsection{Pointing and Guiding}

The driver for the guiding stability is the high photometric precision:
pointing drifts, combined with detector sensitivity variations and possible position-dependent systematics, will introduce apparent position-dependent intensity variations. While significant reduction in the power of such systematics is possible via post-processing (such as in the Kepler mission), it is desirable to keep image drifts at or below the level of the diffraction-limited spatial resolution of the telescope. 
Therefore, a guiding precision of approximately 15mas/10~hr would be targeted.  The Nautilus unit telescopes will use the sun as a coarse attitude reference point and the anti-solar starfield for precise pointing position measurements. The unit telescopes will use the target stars and reference stars within the field of view for fine guiding during long exposure series (typically $\sim$20~hours) before, during, and after planetary transits. 

\subsection{Thermal Management}

At an orbit with an average distance of 1~au from the Sun, the spherical Nautilus units may operate close to room temperature (25~$^\circ$C) with only modest active thermal management (heating). 
The inflatable balloons of the Nautilus units will protect the instrument package (in the interior) from large temperature excursions, as the nitrogen gas and emission/absorption within the balloon redistributes heat. Nevertheless, the high-precision measurements require a thermally stable system (instrument package, lens alignment, and lens itself). Therefore, Nautilus units will actively control the temperature of the elements within the spacecraft and of the MODE lenses. Heating will be provided by battery-powered thermoelectric cells, and excess heat will be dumped at the dark (non-illuminated) side of the Nautilus units, possibly through a metal ring surrounding the MODE lens (as MODE lenses are not exposed to the Sun during normal operations). 

\section{Design, Fabrication, and Scaling}

In this section we review the current status, challenges, and pathways for the optical design and fabrication of large MODE lenses the Nautilus Array concept is based on. This discussion is followed by a summary of the design challenges for the spacecraft architectures. 

\begin{figure*}[!htbp]
\epsscale{1.15}
\plotone{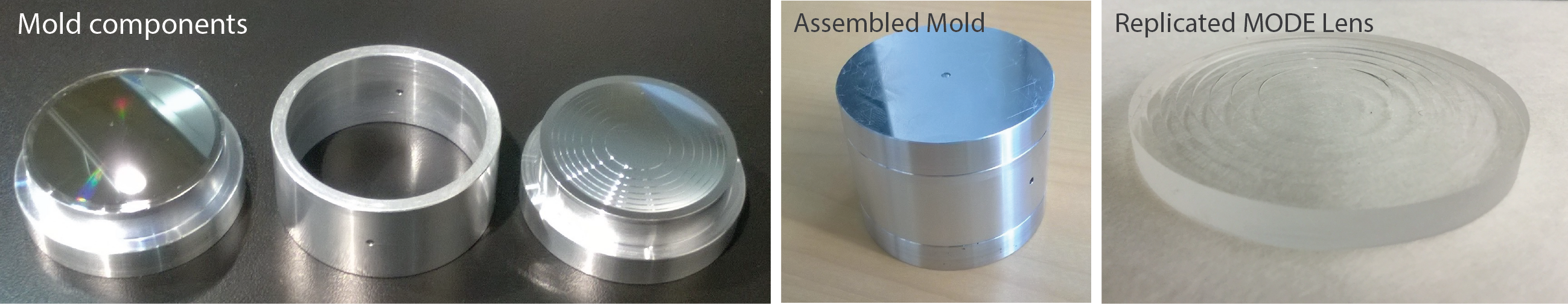}
\caption{Mold-replicated diffractive lenses at the University of Arizona College of Optical Sciences, developed by our team. The mold is produced via diamond milling, the optical design of the MODE lenses provides nearly diffraction-limited image quality over a broad wavelength range with negligible chromatic aberration. The MODE lenses can be readily and very cost-effectively replicated.
\label{MoldingProcess}}
\end{figure*}

\begin{figure*}[!htbp]
\epsscale{1.15}
\plotone{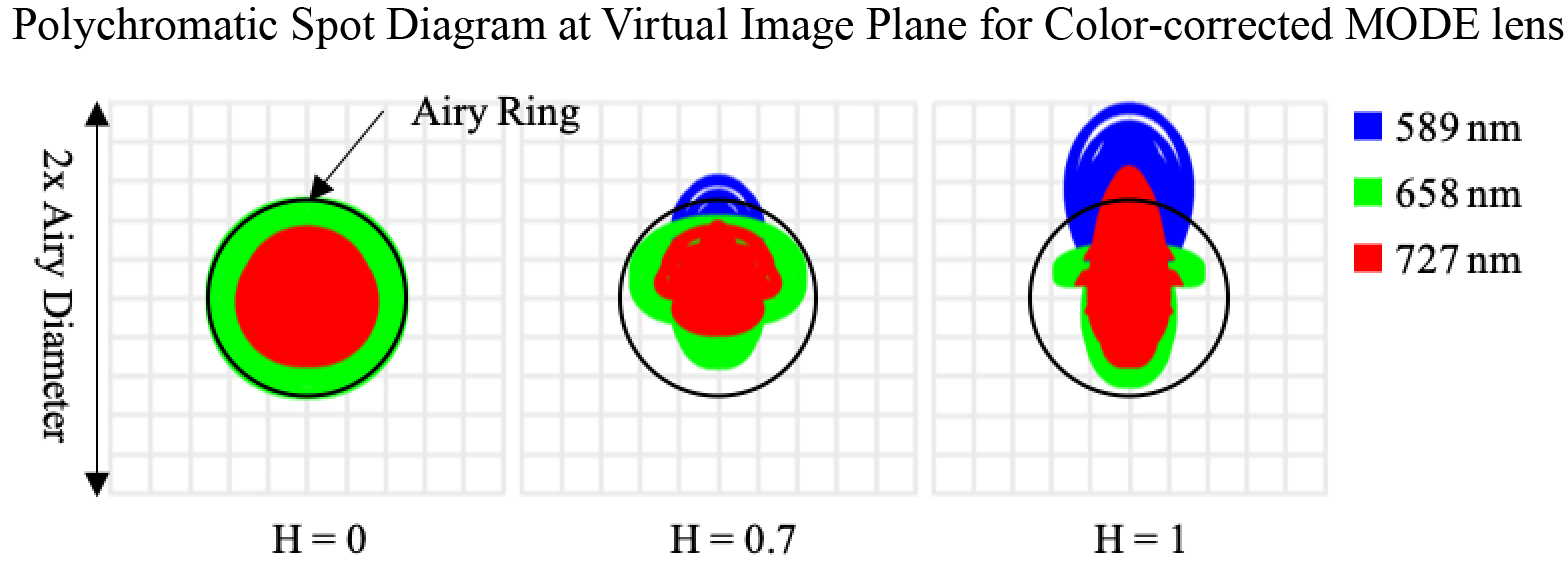}
\caption{Polychromatic spot diagram at the virtual image plane of the color-corrected MODE telescope for an existing 0.24m diameter design. In this design the Airy diameter is 2.94 $\mu$m and the maximum full field of view (H = 1) is 7.5 arcminutes. The image performance is diffraction-limited on-axis (H=0) and the spot size in only slightly larger than the diffraction limit at the edge of the field of view (H=1). \label{SpotDiagram}}
\end{figure*}

\subsection{MODE Lens Fabrication Process}
\label{S:Fabrication}

Due to their non-continuous surface microstructures, MODE lenses cannot be fabricated through traditional grinding and polishing methods. Potential fabrication for non-continuous surfaces include diamond turning, and molding, gray-scale lithography, deep-reactive ion etching, UV imprinting, and glass slumping. Among these, diamond turning and molding are the most powerful approaches for MODE lens fabrication due to their accuracy and scalability. The combination of diamond turning and pressure molding offers very powerful and flexible fabrication paths for MODE lenses: diamond turning enables precise fabrication and pressure molding enables reliable and low-cost replication.

Ultra-precision diamond-turning machines have been successfully used to fabricate conventional lenses as well as diffractive optical elements \citep[e.g.,][]{LeeCheung2003,HuangLiang2015}. For example, state-of-the-art Moore Nanotech 350FG freeform generators are capable of diamond turning or milling  MODE lenses or MODE lens segments with diameters up to 0.6~m. Precision glass compression molding is a replicative process that allows the production of high-precision optical components from glass and polymer \citep[][]{ZhangLiu2017}, including those of diffraction surfaces \citep[][]{Huang2013,Nelson2015}. By using chalcogenide glasses precision molding also allows replicating optical elements for infrared applications \citep[][]{Staasmeyer2016}. Compression molding has been successfully used to mold glass freeform optics from diamond-machined molds \citep[][]{He2014}.

By combining diamond turning/milling to fabricate molds and glass press molding, it is thought to be possible to replicate large-aperture MODE lenses. Our team is actively developing this technology and fabricated, replicated, and tested lens prototypes. To illustrate the fabrication approach, we diamond-turned molds and molded the MODE lens from poly(methyl methacrylate). One of such prototype is shown in Figure~\ref{MoldingProcess}).  The measured lens profile and image quality verified the molded surface shape and quality. Our University of Arizona-based team is developing this technology toward very large-aperture MODE lenses that can be replicated reliably and at low cost.

\subsection{MODE Lens Prototypes}

As part of the MODE lens development effort at The University of Arizona our team has designed and fabricated several generations of diffraction-based lenses, from single-order to more complex (M=1,000) MODE lenses. The latest prototype MODE lens combined a high-order diffractive lens (M=1,000) at the front surface with a single-order Fresnel lens on its back surface. The optical design has been optimized to provide diffraction-limited performance in the astronomical R-band (589\,nm to 727\,nm). Laboratory optical tests with super-continuum laser and imaging tests demonstrated that the measured performance is consistent with that predicted by the physical optical models (Milster et al., in prep). Prototypes equipped with additional, conventional color-corrector optics (much smaller in diameter than the MODE lens) are predicted to provide diffraction-limited performance over broad wavelength range ($>$200\,nm).

\subsection{Optical Quality Assessment}
\label{S:Metrology}

In order to manufacture, align, and assemble a large MODE lens-based space telescope system, specialized metrology concepts and solutions testing and verifying its optical performance need to be developed and applied. The challenging science goals of next-generation space telescopes are often achieved through a new optical design  and components, such as MODE lens. Measuring and aligning those complex optical surfaces and components require large dynamic range in metrology, high accuracy, comprehensive spatial-frequency coverage, and real-time data acquisition and analysis to test the multi-scale optical features. Unlike the traditional small size diffractive optical component, the diffractive optical surface of the MODE lens has the precision microscopic surface profile over its large aperture area, enabling its light collecting power. As stray light and surface scattering must be controlled for high-quality imaging performance, the nano-scale surface roughness must be measured and controlled. In order to realize the next generation optical systems producing a near-optimal point spread function (PSF) with superb imaging quality, an entire spectrum of the optical system's wavefront must be measured and confirmed during the telescope manufacturing, assembly, and testing process. 

As the entire spatial frequency spectrum of the optical surface/wavefront errors has to be controlled, the MODE-based space telescope optics need to be modeled and specified using a power spectral density (PSD) or structure function  \citep[][]{Hvisc2007,Parks2010}. \citet[][]{Parks2008} experimentally demonstrates the severe image quality degradation due to the presence of mid-to-high spatial frequency surface errors.

Phase shifting deflectometry is applied to measure, align, assemble and evaluate the performance of a large optical system with nanometer level accuracy through its direct slope measuring capability  \citep[][]{Su2010,Oh2016}. During fabrication the quality of the MODE-lens local surface finish (i.e., micro roughness RMS value) is monitored and sampled across the large aperture using portable white light interferometer  \citep[][]{Parks2011}. Various metrology systems covering different range of spatial frequencies measure and test the MODE-based space telescope by providing a comprehensive PSD evaluation similar to the 4.2 m Zerodur primary mirror of Daniel K. Inouye Solar Telescope (DKIST) tested with a suite of metrology systems as shown in 
\citep[][]{Kim2016}.  

Other key components of the space telescope system utilize aspheric optics to achieve better achromatic imaging performance for a larger field of view within a compact and light-weighted design. For most cases, temporal phase-shifting interferometry using a null component such as Computer Generated Holograms that provides high accuracy wavefront/surface measurement data with sufficient spatial frequency sampling. During the manufacturing and aligning process, in order to guide the processes, high dynamic range metrology methods are utilized. The wide range ensures the measurement of the optical component's or system's quality when it is still far away from its final specification performance. For instance, the 4--8 m diameter class large precision optics manufacturing process and final testing of the Giant Magellan Telescope (GMT) and DKIST primary mirrors were guided with a successful rapid convergence by utilizing non-null deflectometry measurement feedback \citep[][]{Su2012,Huang2015}.

For the alignment of MODE lens systems (as also applicable for active/adaptive wavefront correction or bending mode measurements) instantaneous metrology is applied using a multiplexed deflectometry solution. The real-time metrology system uses a multiplexed color fringe pattern in $x$ and $y$ spatial frequency domain in order to contain and process six fringe patterns from a single-shot data acquisition offering about 25 nm RMS accuracy  \citep[][]{Trumper2018}. Such a dynamic measurement and active characterization of the MODE lens telescope system will monitor and verify the opto-mechanical performance as a function time, orientation with respect to the gravity, and thermal gradient changes. 
An overview of the baseline MODE lens optical specification as a function of spatial frequency (i.e., cycles per aperture) is given in Table~\ref{T:OptSpec}.

\begin{deluxetable*}{p{4cm}p{3cm}p{2.5cm}p{5cm}}
  \caption{Overview of the spatial frequency optical specification for the MODE lens at wavelength $\lambda$. \label{T:OptSpec}}
\tablehead{\colhead{Specification}&\colhead{Spatial Frequency}&\colhead{Requirements} &  \colhead{Note}} 
\startdata 
Transmitted wavefront error due to the surface figure error &
Low ($<$4 cycles per aperture) &$<\sim0.017\lambda$ RMS &
Interferometric measurement using HeNe laser ($\lambda$ = 633nm)\\
Transmitted wavefront error due to the mid-to-high spatial frequency errors&
Medium (4--60 cycles per aperture) &
$<\sim0.015\lambda$ RMS &
Combining interferometric and deflectometric measurements \\
Transmitted wavefront error due to the mid-to-high spatial frequency errors &
High ($>60$ cycles per aperture) &
$<\sim0.017\lambda$ RMS &
Combining interferometric and deflectometric measurements\\
Surface roughness error due to the micro surface finish &
Very high (sampling resolution $<100 \mu m$/cycle)&
$<\sim2~nm$ RMS &
Measurement using whitelight interferometer over 1$\times$1~mm area with 500$\times$500 sampling points\\
\enddata
\end{deluxetable*}

\subsection{Optical Performance}

{\changed As an indication of system {\changed optical} performance, spot diagrams are calculated that show geometrical ray intercepts at the image plane{\changed. These diagrams} trace bundles of rays through the optical system from a point source {\changed(like a star) at a large distance.}  Figure~\ref{SpotDiagram} shows spot diagrams of a 0.24m prototype MODE {\changed lens} telescope at an intermediate image plane after color correction. {\changed The location of the idealized 1st Airy ring minimum is shown as a dark circle}. The Airy {\changed pattern} is the image-plane light distribution due to diffraction of the optical system illuminated by a distance point source.  The diameter of the Airy ring is dependent on the wavelength and $f$ number. In this case, the Airy ring is calculated from the prototype system parameters with a central wavelength of $\lambda_c=658$~nm and an $f$-number of 1.83. The image is considered diffraction limited if all of the geometrical ray intercepts fall within or near the Airy ring. In Figure~\ref{SpotDiagram}, the geometrical ray intercepts are calculated from three wavelengths of the point source, 589~nm, 658~nm, and 727~nm, which are shown as blue, green, and red colors, respectively.  Calculations are also made from three field angles of the point object, from on-axis (at H = 0) to a maximum field angle of 7.5 arcminutes (H = 1).  The prototype design is diffraction-limited on axis at H = 0 and out to 70\% of the maximum field of view (H = 0.7).  It is nearly diffraction-limited at the maximum field of view (H = 1).  Degradation of the system performance with increasing field {\changed angle} is due to residual wavelength-dependent aberrations. While this design shows a 0.24m-diameter MODE-lens based telescope system's performance, similarly diffraction-limited performance can be achieved by larger systems. }

\section{Discussion}  
\label{S:Discussion}
  The Nautilus mission concept described here envisions a space telescope array based on low-cost, replicated, 8.5m-diameter inflatable space telescopes utilizing novel, ultralight diffractive optics. Nautilus and the enabling technology will transform the design, construction, operation, and launch of space telescopes for scientific, commercial, and other applications. In the following we briefly compare MODE lens-based technology and, specifically, the Nautilus design to the state of the art in space telescopes, and review the primary advantages of the MODE technology over current mirror-based approaches.
  
\subsection{Comparison of Capabilities to the State of the Art}  
No existing telescope is capable of searching for atmospheric biosignatures in exoplanets. JWST may be able to search for water and methane in the most favorable transiting exoplanets around very nearby red dwarf host stars, but probably in not more than 2-4 habitable zone earth-sized planets. The mission concepts HabEx and LUVOIR would utilize high-contrast direct imaging to study earth-like planets; {\changed they may image 50--300 planetary systems and  search for biosignatures in up to about 10 to $\sim$60 habitable planet candidates}, respectively. 

In contrast, Nautilus is a system concept developed to survey $\sim$1,000 earth-like 
planets, providing more than an order-of-magnitude increase over the capabilities of even the most ambitious missions yet studied. The proposed survey can be accomplished with an array of telescopes which combine light incoherently. Achieving this level of light collection in a single, phased aperture is not a realistic possibility for the foreseeable future.

In terms of light-gathering power Nautilus offers orders-of-magnitude increase over current facilities. Currently the largest diameter space telescopes are HST ($D=$2.4~m), and Herschel ($D=$3.5m), with the $D=$6.5m JWST soon to follow. A single element of the Nautilus concept will exceed HST’s light-gathering power by a factor of 12.5 and the Nautilus array will exceed JWST’s collecting area by a factor of 60 \citet[][]{LUVOIR2018}.  

\subsection{Tolerance to misalignments}
\label{S:Misalignments}

Unlike reflecting telescopes, our MODE lens design is inherently more tolerant to optical element misalignments, a fact that will significantly reduce fielding costs (e.g., \citealt{LoArenberg2006}). If a mirror orientation is tilted by angle $\alpha$, the reflecting beam will be walk off by $2\alpha$, which requires a very tight alignment tolerance and complex control solutions.  However, with a basic first-order geometrical optics analysis, a chief ray going through the center of a {\em refractive} lens does not change its direction although the lens is tilted. In a similar manner, transmissive refractive/diffractive optics are insensitive to surface figure errors including mid-to-high spatial frequency errors. For example, an anomaly with height $h$ on a mirror surface in space will induce $2h$ OPL (Optical Path Length) change due to its double-path nature. However, for a lens with refractive index of $n$, the same surface anomaly will cause only $(n-1)\times h$ OPL difference, i.e., only $\sim$0.5\,h change in OPL (assuming a typical $n = 1.5$). Also, if the thin MODE lens is bending or locally rippling while it is maintaining the thickness of the MODE lens, there is almost no OPL change since the front and back surfaces are moving together. This robustness of the alignment and shape error tolerance is one of the most fundamental strengths of the MODE lens based telescope system.

\subsection{Mass Comparison to Mirrors}
\label{S:MassComparison}

As an illustration of the anticipated mass advantage of MODE-based space telescopes over reflecting space telescope, we contrast scaled-up versions of the HST and JWST mirror systems with MODE telescopes.
HST uses a 2.4m-diameter, 33 cm thick, monolithic mirror, whose weight is reduced with respect to a conventional monolithic mirror by about a factor of five through the implementation of a honeycomb structure in the body of the mirror. The mass of HST's mirror is approximately 826~kg, which corresponds to about 7.4\% of the total observatory mass. The HST mirror provides an excellent reference point for relatively light-weight monolithic space telescope mirrors. 

JWST is a primarily infrared telescope with diffraction-limited optical performance at and beyond 2 microns. It utilizes 18 gold-coated beryllium mirror segments, each of which are 20.1~kg. The segments are periodically co-phased between observations. Considering the mirror control structure (wire harness), the complete primary segment assembly for each mirror segment is 39.48~kg. This leads to a combined mass of 710 kg for the 6.5\,m mirror, corresponding to 9.6\% of the total observatory mass. JWST is the natural reference for state-of-the-art segmented space telescope mirrors.

In Table~\ref{T:MassComparison} we compare the masses of hypothetical 8.5-m diameter mirrors that use HST- and JWST-like mirror systems. We provide two bracketing cases for the mass scaling with diameter: an optimistic case when the mass $M$ is directly proportional to the $D$ collecting area of the mirror ($M\propto D^{2.0}$, and a more conservative one ($M\propto D^{2.8}$), in which the larger area also translates into thicker mirror (or additional co-phasing system). 
We compare these extrapolated mirror masses to two MODE lenses: a thin ($h=0.5$~cm thick, optimistic case) and a thick ($h=5$~cm thick, pessimistic case) lens. We note here that the thickness of the MODE lens will be likely set by mechanical structural considerations and not the optical design, as even few mm thick MODE lenses can provide excellent image quality.
To calculate the mass of the MODE lens (see \S\,\ref{LensWeight})we consider its volume ($R^2 \pi \times h$), the volume-filling factor $\phi$=0.5, and assume a glass density of $\rho$=2,500 kg/m$^3$ . 

Table~\ref{T:MassComparison} demonstrates that the MODE lens technology is expected to be a factor of 2 (worst case) to 70 (best case) lighter than an HST-like honeycomb mirror. Compared to JWST-like segmented mirrors, MODE lenses have similar mass (between 3 times lighter to about three times heavier). Compared to a conventional mirror (about 5 times more massive than honeycomb mirrors), MODE lenses would provide about two orders of magnitude lower mass.  

Not only will MODE technology enable ultralight light-collecting capability (on par or better than the lightest mirrors), further significant reduction in the total mass for a MODE-based telescope system is expected given the overall lighter support structure required and due to the fact that the MODE lenses are much more tolerant to misalignments.

\begin{deluxetable*}{lcccc}
  \caption{Mass comparison between different mirror and MODE lens primaries. Current and Nautilus-like diameters listed. For the 8.5m diameter HST and JWST-like mirror systems are scaled with diameter. The mass of the MODE lens is calculated from its volume and typical glass density (2,500 kg/m$^3$). Two different thicknesses are shown for the MODE lens to illustrate anticipated mass range. MODE lenses have the potential to provide low-cost and very low-weight alternatives to mirror systems, with extrapolated mass well below HST-like honeycomb mirrors and similar to JWST's ultralight segmented mirror system. \label{T:MassComparison}}
  \tablehead{\colhead{Case}&\colhead{HST Honeycomb}&\colhead{JWST Segmented}&\colhead{MODE 0.5 cm}&\colhead{MODE 5 cm}} 
  \startdata 
Current Design & 2.4m, 826 kg & 6.5m, 710 kg & --  \\
\hline
\multicolumn{4}{l}{\em Assuming $M\propto D^{2.0}$ scaling for HST and JWST:} \\
Scaled to 8.5m & 10,360 kg & 1,214 kg & 425 kg & 4,255 kg  \\
\hline
\multicolumn{4}{l}{\em Assuming $M\propto D^{2.8}$ scaling for HST and JWST:} \\
Scaled to 8.5m & 28,494 kg & 1,505 kg & 425 kg & 4,255 kg  \\
\enddata
\end{deluxetable*}
  
\subsection{Potential for lower launch costs} 
\label{S:LaunchCosts}

The Nautilus concept benefits from {\changed  a potential for} greatly reduced launch costs through four factors: Firstly, MODE lens-based telescopes will be much lighter than telescopes based on monolithic mirrors and about as light as the lightest segmented mirror systems (\S\,\ref{S:MassComparison}). Secondly, with the typically two orders-of-magnitude more relaxed alignment tolerances (\S\,\ref{S:Misalignments}), the structural support requirements are milder and allow for the use of light-weight structural elements, such an inflatable deployment mechanisms (\S\,\ref{S:NautilusArray}). The light-weight and inflatable structural elements represent significant further reductions in the telescope's mass. Thirdly, MODE lens systems can provide simultaneously fast systems (small focal ratios) {\em and} wide field of view. This allows the MODE telescopes to be very compact (f/1.0 systems), thus alleviating the need for light path folding and secondary mirrors. Fourthly, the very compact launch configuration and low weight enables the simultaneous launch of many unit telescopes in a single launch fairing (e.g., up to 15 with SpaceX/BFR or 25 with NASA SLS B2 Long). This dramatically reduces the per-telescope launch costs. 
{\changed T}he Nautilus concept is too preliminary to allow for reliable costing{\changed. It seems, however, that} due to the four factors discussed above {\changed MODE-lens based systems -- and, in particular, the Nautilus Observatory -- has potential to provide significantly lower launch cost solution} than those following more conventional design.

\subsection{Potential for lower mission costs and risks} 

 It is argued that current mission costs and complexity are driven by the so-called ``space spiral'', in which higher reliability requires longer development phase, which results in fewer missions and higher mission costs, which -- in turn -- requires even higher reliability  \citep[][]{NewSMAD}. This self-reinforcing cycle arguably drove mission costs and it is estimated that at present day the average cost of space systems launched by the US is about 3 billion USD per launch \citep[][]{NewSMAD}. 
 
 The Nautilus Observatory represents a new approach to space telescope fabrication, one in line with the development of small satellites and diversified launch capabilities. 
{\changed The Nautilus concept has the potential to reduce mission costs in three major ways}: Firstly, it utilizes a low-cost (replicated) optical element instead of massive and complex mirrors systems. 
 This reduces the fabrication cost of one of primary components of telescopes that conventionally drives mass, risk, and cost budgets. This represents a fundamental paradigm shift, making the production of space telescopes far more economical.
 Secondly, the MODE lenses provide ultralight-weight alternatives to mirrors and enable light-weight telescope structures also utilizing inflatable elements, translating into major reduction in launch costs (\S\,\ref{S:LaunchCosts}). 
 Thirdly, unlike many past space telescopes and space observatories, Nautilus is envisioned not as a unique, high-reliability, and very expensive telescope, but an array of replicated, identical, relatively low-cost telescopes. This major difference is enabled by the ability to efficiently replicate the key optical element by using the MODE technology.
 
  The model in which many unit telescopes are built and launched also alleviates the very high reliability requirement: compromised operation (or even failure) of one unit telescope would not compromise the array's overall capabilities. Similarly, the instruments envisioned for Nautilus units are simple and replicated; less capable than typical HST and JWST instruments, but, also with individually relaxed fault tolerance, can be built for a fraction of the cost. In addition, the risks could be better distributed than possible in the current single, unique mission model: we can envision the launch of one or two smaller-size demonstrator units to mitigate risks.

{\changed The specific cost of the MODE lens and its associated system cannot be rigorously derived at this early TRL and mission concept maturity level (CML).  However, the technology offers a likely cost advantage over traditional systems and a qualitative argument can be provided to establish that MODE lenses are less expensive than traditional approaches.  Let the cost of each stage $i$ in the development of the MODE and traditional optics be denoted M$_i$ and T$_i$, respectively.  Correspondingly, the total costs of a MODE-based system is $\Sigma_i M_i$, while the total cost of a traditional mirror-based system is $\Sigma_i T_i$.

Since at this early system TRL and CML we cannot reliably predict total cost, we use another method of analysis. This method is a heuristic one, and shows that $M_i \leq T_i$
for all $i$. Consider ~\ref{T:Costs}, which lists the steps in realized an optic and the assessment at each stage of why traditional approach is likely to be equal or more costly than the MODE technology.

\begin{deluxetable*}{p{4.5cm}p{7cm}p{2cm}}
  \caption{Assessment of Elements of Cost{\changed}. \label{T:Costs}}
\tablehead{\colhead{Element of Cost}&\colhead{Traditional Approach}&\colhead{MODE Relative Cost}} 
\startdata
Design & Same   & Same \\
Materials & Larger due to the greater mass of materials. Structure more costly than & Lower \\
Manufacturing Tooling & Larger due to the need for more types of machines & Lower\\
Manufacturing Recurring Cost & Larger due to greater time to produce and optic, each steps is longer in time than the MODE molding step  & Lower \\
{Alignment, Integration  \& Testing} &  Larger due to need to integrate mirrors on to structure
 & Lower \\
Verification & Same  & Same \\
\enddata
\end{deluxetable*}

As shown in Table~\ref{T:Costs}, our assessment is that for each step, the cost for the MODE lens is less than or equal to the cost of traditional methods, meeting the condition that $\Sigma_i M_i < \Sigma_i T_i$ and, therefore, indicating the cost efficiency of the MODE technology. As the technology and concept develops we will be able to improve upon the originally heuristic cost argument more quantitative and precise, but the qualitative result given is likely not to change.
}

\subsection{Scalability and Scalability Challenges}
Due to its relatively low production and launch costs and the identical multi-spacecraft model that is relatively new to astrophysical space telescopes, the general Nautilus system proposed here provides an easily scalable approach. Such multi-spacecraft model (multiple identical units) are used commercially (Iridium system) and for geo- and planetary sciences (Voyagers, Mariners, Mars exploration rovers, etc.) to reduce per-unit costs and risks and to extend capabilities.
Furthermore, telescopes utilizing similar architecture, but increasing in size, could demonstrate feasibility and mitigate risks, while producing scientific data. 

As discussed in Section \ref{S:Fabrication} the combination of optical high-precision diamond-turning and compression molding has a clear potential for enabling the efficient replication of very large-scale diffractive optical elements. In Section~\ref{S:Metrology} we reviewed considerations for optical quality monitoring for such large-scale optical elements and showed that the required technology already exists today and only minor changes will be required to adopt it for MODE lenses.

Here we will briefly review three challenges that must be overcome to enable the replication of large-scale MODE lenses. First, diamond turning and molding machines must be scaled up significantly. Both of these technologies are fundamentally not very sensitive to spatial scales, i.e., the construction of large diamond turning and molding machines is thought to be entirely possible by just building larger versions of the current machines. No change in technology is required. 

Second, the structural integrity of the large-diameter MODE lenses must be preserved during launch (unless molded in space). Large-scale, relatively strong, yet thin transmissive glass elements exists in a variety of fields. For example, car windshields (layered glass panels) are large-diameter and strong, yet typically only 4-6 mm thick. Nevertheless, the fabrication, handling, and launch of very large, light-weight optical glass elements will clearly represent a challenge. Third, the deployment of the large MODE lenses through a light-weight structural element (such as a balloon, \S\,\ref{S:NautilusArray}) must be demonstrated. Inflatables and optical deployables have a long heritage in satellites (e.g., ECHO-1, or solar panels), and heritage solutions may already exists for smaller scales. The fact that MODE lenses are very tolerant to misalignments (deployment errors) is encouraging. Although there are reasons to believe that all three of these challenges are surmountable in the very near future, MODE lens technology development and mission concept design must mitigate these risks.

\subsection{Science Impact}

MODE lens-based, very large-aperture telescopes in general, and the Nautilus concept specifically, offer a possible revolution in the light-gathering power in the astronomical space telescopes. The greatly enhanced light-gathering power equals greatly enhanced sensitivity to faint astrophysical objects (such as the earliest, very high redshift galaxies; supernovae at high redshifts; individual stars in nearby resolved galaxies, or small minor bodies in the Solar System). The enhanced sensitivity also enables more precise and higher-cadence time-resolved observations. One important application of such observations is the characterization of transiting extrasolar planets, the science case that motivated the Nautilus concept described in this manuscript. 
As demonstrated, our baseline Nautilus concept may enable spectroscopic studies of approximately one thousand potentially habitable Earth-sized exoplanets. Such a survey would undoubtedly revolutionize astrophysics, planetary sciences, and astrobiology. The spectroscopic observations would enable the identification of several key atmospheric absorbers and would provide a pathway to determine or constraint the atmospheric composition. The survey envisioned here is distinct from other surveys proposed or planned due to its very large sample size: the ability to study a thousand Earth-sized habitable zone planets may likely be essential for understanding the complexity and diversity of extrasolar planets \citep[e.g.,][]{Seager2014,Apai2017SAG15,Bean2017}. 

A sample size of a thousand planets would allow, for example, identification of potential trends between atmospheric absorbers and bulk properties of the planets. Comparison of the planets' atmospheric composition to the stellar irradiation received may allow empirical mapping of the inner and outer boundaries of the habitable zone \citep[][]{Kopparapu2013,Kopparapu2018}, and identification of possible atmospheric loss mechanisms \citep[e.g.,][]{OwenWu2016}. \citet{Seager2014} argues that a sample size of 1,000 or greater potentially Earth-like is likely required for a confident, statistical identification of life-bearing planets.

\subsection{Possible Pathway toward Large Diffractive Telescopes}

{\changed The fundamental {\changed theme of this paper is to explore the conjecture that} large astronomical telescopes could be built based on multi-order diffractive engineered lenses and that such telescopes could be uniquely well-suited to address astrophysical problems that require large light-collecting area. The Nautilus Observatory concept described {\changed herein} is not a complete design reference mission and it is not informed by detailed trade studies; instead, our study shows that possible solutions exist to most challenges MODE{\changed}-lens based telescope architectures may pose. 

The technology readiness level of several components of the Nautilus concept is low; chief among these are the MODE lenses themselves {\changed with} TRL2-3. In order to verify and realize the potential for very large {\changed MODE} lenses for astronomical observations, a significant technology development and demonstration program is required. We briefly describe here a possible {\changed technology maturation} pathway toward large MODE-lens based telescopes. 

First, technology development is required to demonstrate that high-quality MODE lenses can be fabricated and replicated {\changed with} sub-meter diameters. Second, these lenses must be demonstrated in astronomical observations, firmly establishing {\changed such} lenses at TRL3. Operational demonstration in thermal-vacuum chambers will help move MODE {\changed lens} technology to TRL5. A parallel technology development effort is required to scale up the fabrication/replication technology by building larger free-form optical fabrication and molding machines, preferably to 1--3m diameters, and possibly beyond. With MODE lenses at TRL4--6, small-scale pathfinder, science-driven space missions will become viable, including small satellites and stratospheric balloons. 

As MODE technology {\changed addresses a fundamental attribute of space telescopes (light collection), interesting science cases exist for even relatively small MODE-lens }based space telescopes. After a successful demonstration of the MODE lens technology in space or near-space environments on a SmallSat or stratospheric flight, a pair of 1--1.5m diameter units may be flown as a NASA Small Explorer or Mid-Explorer mission, or a single 8m-diameter telescope could be flown as a Probe-class mission. If successful, a scaled-up and replicated version of these unit telescopes could serve as the first step in realizing the Nautilus Observatory or an observatory with a similar scope. 

We note here that the initial steps of this process are underway: small (0.05m diameter) MODE lenses have been fabricated, replicated, and demonstrated already; our team is currently working toward developing 0.24m diameter MODE lens-based telescopes and their on-sky demonstration is scheduled for Winter 2020. In addition, we are planning small satellite and balloon-borne MODE telescopes.
}

\section{Conclusions}

One of the most fundamental properties of astronomical telescopes is their light-gathering power; yet, increase in mirror diameter over the past two centuries has been slow compared to performance increases seen in complementary fields (detectors, engineered materials, computer processors). In this study we described a very large astronomical telescope based on a novel, {\changed MODE} lens design{\changed ,} and an ultralight, inflatable spacecraft. {\changed Our concept focuses on the unique aspects of MODE-based telescopes and it is not a complete, optimized mission concept.} The {\changed notional} Nautilus telescope concept introduced here is motivated by the science goal of surveying one thousand Earth-sized, potentially habitable exoplanets -- a study that is important to understand the diversity of Earth-like planets, but requires light-gathering power far beyond projected capabilities. 

The key results of our study are as follow:\\
{\em 1)} Multi-order diffractive lenses provide ultralight and very large diameter alternatives to astronomical reflectors.\\
{\em 2)} MODE lenses potentially offer three key advantages over telescope mirrors: Much lower weight per unit area, less sensitivity to misalignments/deformations, and efficient replicability through optical molding processes.\\
{\em 3)} We describe a {\changed novel and notional} telescope concept that will allow transmission spectroscopy of one thousand transiting, Earth-sized, potentially habitable planets at visual/near-infrared wavelengths.\\
{\em 4)} We evaluate four different target selection criteria for the exoplanet host stars (different spectral types) and assess the distances up to which a telescope must be capable of probing atmospheric biosignatures to search for life in 1,000 earth-sized habitable zone planets. \\
{\em 5)} We use the Planetary Spectrum Generator to calculate the expected transit spectra for some the best-case and worst-case targets. We find that a 35x8.5m array of telescopes is sufficient to probe biosignatures in 1,000 transiting Earth-sized habitable zone exoplanets. \\
{\em 6)} The Nautilus concept is based on a large array of identical unit telescopes, each equipped with a 2.5~m diameter lens optimized for wide-field imaging and exoplanet transit searches, and with an 8.5\,m-diameter MODE lens optimized for high-precision, moderate-resolution transit spectroscopy.\\
{\em 7)} Individual units can be used for wide-field surveys or targeted exoplanet transit searches, while the array -- through the non-coherent combination of the light intensity signal from multiple units  -- enables the detection of faint light sources (e.g., very high redshift galaxies) as well as low-amplitude time-varying signal (e.g., exoplanet transit spectroscopy).\\
{\em 8)} The costs of the array are minimized by utilizing MODE lenses replicated through molding, by equipping each telescope with simple and identical instruments, and by launching 15 unit telescopes in a single launch.\\
{\em 9)} With two launches of next-generation rockets (SpaceX/BFG or NASA SLS B2) enough unit telescopes can be launched for the Nautilus Telescope Array to provide a light-gathering power equivalent to a 50\,m diameter space telescope. \\
{\em 10)} Although diffractive optical elements have flown as part of space instruments and small-scale MODE lenses exist, significant technology development is necessary before truly large-aperture MODE telescope could be built. We discuss the key technology development challenges for MODE telescopes.\\

In summary, the concept described here offers a pathway to break away from the cost and risk growth curves defined currently by mirror technology, and has the potential to enable very large and very light-weight, replicable technology for space telescopes. An example application of the Nautilus concept promises a revolutionary atmospheric biosignature survey of a thousand potentially Earth-like exoplanet.


\acknowledgments
The authors thank the anonymous referee whose comments have improved the manuscript. We acknowledge helpful discussions and input from P. Apai, P. Atcheson, J. Breckinridge, B. Crill, J. M. Grunsfeld, B. T. Jannuzi, M. Marley, D. S. Lauretta, E. Mamajek, B. V. Rackham, N. Siegler, Z. Wang. The authors are particularly grateful for other members of the Nautilus team who contributed to the prototype MODE lens design, fabrication, and tests (O. Spires, Y-S. Kim) and project management (C. Fellows). This study is funded in part by the Gordon and Betty Moore Foundation. We acknowledge A. Conti and the Northrop-Grumman Advanced Systems for a workshop on the future of exoplanet exploration and space telescopes, which seeded the current study. DA also acknowledges the NASA Exoplanet Exploration Program Office and the broader EXOPAG community for the development of the scientific and programmatic context that enabled this study. DA is grateful for the Max Planck Institute for Astronomy, Heidelberg, for support and hospitality during a sabbatical visit during this study. The results reported herein benefited from collaborations and/or information exchange within NASA's Nexus for Exoplanet System Science (NExSS) research coordination network sponsored by NASA's Science Mission Directorate.

Author contributions: DA initiated the Nautilus project, lead the mission concept definition, the development of the science case, created many of the figures, and drafted the manuscript. TM led the optical design, the invention of the MODE technology, and contributed sections on the optical design. DWK contributed sections on optical metrology and contributed to the conceptual design of the Nautilus system. AB led the transit simulations and the observation time assessment. GS contributed to the manuscript and to the definition of the science case. RL contributed to the optical fabrication section of the manuscript and leads optical fabrication of the MODE lens prototypes. JA contributed ideas to the Nautilus spacecraft concept and led the mission cost discussion. 

%

\vspace{5mm}
\facilities{}


\software{Python:NumPy/AstroPy}

\bibliographystyle{aasjournal}
\bibliography{nautilus_apai}

\appendix

\begin{deluxetable}{ccll}
  \caption{Assumed parameters for sample size definition. \label{Table:SampleAssumptions}}
  \tablehead{\colhead{Parameter}&\colhead{Value}&\colhead{Description} & \colhead{Refs.} } 
  \startdata 
$R_*(F)$ & 1.30 & Stellar Radius for F7V-type star & \citet{PecautMamajek2013} \\
$R_*(G)$ & 0.95 & Stellar Radius for G7V-type star & \citet{PecautMamajek2013} \\
$R_*(K)$ & 0.65 & Stellar Radius for K7V-type star &\citet{PecautMamajek2013} \\
$R_*(M)$ & 0.12 & Stellar Radius for M6.5V-type star & \citet{PecautMamajek2013}\\
\hline
$M_*(F)$ & 1.21 & Stellar Mass for F7V-type star & \citet{PecautMamajek2013} \\
$M_*(G)$ & 0.96 & Stellar Mass for G7V-type star & \citet{PecautMamajek2013} \\
$M_*(K)$ & 0.65 & Stellar Mass for K7V-type star &\citet{PecautMamajek2013} \\
$M_*(M)$ & 0.10 & Stellar Mass for M6.5V-type star &\citet{PecautMamajek2013} \\
\hline
$L_*(F)$ & 0.36 & Log. Stellar Luminosity for F7V-type star & \citet{PecautMamajek2013} \\
$L_*(G)$ & -0.12 & Log. Stellar Luminosity for G7V-type star & \citet{PecautMamajek2013} \\
$L_*(K)$ & -0.98 & Log. Stellar Luminosity for K7V-type star & \citet{PecautMamajek2013}\\
$L_*(M)$ & -3.09 & Log. Stellar Luminosity for M6.5V-type star & \citet{PecautMamajek2013}\\
\hline
$\eta_\oplus(F)$ & 0.3 & Occurrence rate of Hab. Zone Earth-size planet, F7V-type host star & Based on SAG13 Meta-study\footnote{https://exoplanets.nasa.gov/exep/exopag/sag/} \\
$\eta_\oplus(G)$ & 0.3 & Occurrence rate of Hab. Zone Earth-size planet, G7V-type host star & Based on SAG13 Meta-study \\
$\eta_\oplus(K)$ & 0.6 & Occurrence rate of Hab. Zone Earth-size planet, K7V-type host star & Based on SAG13 Meta-study\\
$\eta_\oplus(M)$ & 1.5 & Occurrence rate of Hab. Zone Earth-size planet, M6.5V-type host star & Based on SAG13 Meta-study\\
\hline
$N_{10pc}(F)$ & 6 & Number of F stars within 10~pc & \citet{Henry2018}  \\
$N_{10pc}(G)$ & 20 & Number of G stars within 10~pc  & \citet{Henry2018} \\
$N_{10pc}(K)$ & 44 & Number of K stars within 10~pc  &\citet{Henry2018} \\
$N_{10pc}(M)$ & 248 & Number of M stars within 10~pc  & \citet{Henry2018}\\
\hline
\enddata
\end{deluxetable}
\end{document}